\documentclass[aps,prb, twocolumn]{revtex4}
\usepackage{amsmath}
\usepackage{amsfonts}
\usepackage{amssymb}
\usepackage{dcolumn}
\usepackage{bm}
\usepackage{graphicx}
\usepackage{epstopdf}
\usepackage{color}

\def\BaOsS {Ba$_2$NaOsO$_6$ }
\def\BaOs  {Ba$_2$NaOsO$_6$}
  \def \Na {$^{23}$Na }
   \def \NaE {$^{23}$Na}

    \def \dqS {$\delta_q$ }
   \def \dq {$\delta_q$}
   \def \ie {{\it i.e.} }

\begin{document}

\title{Nature of Lattice Distortions in Cubic Double-Perovskite Ba$_2$NaOsO$_6$} 

\author{W. Liu$^{1}$, A. P. Reyes$^{2}$, I. R. Fisher$^{3,4}$, and V. F. Mitrovi{\'c}$^{1, \dag}$}
%\affiliation{}
\address{$^{1}$Department of Physics, Brown University, Providence, RI 02912, U.S.A.\\
$^{2}$National High Magnetic Field Laboratory, Tallahassee, FL 32310, USA\\
 $^{3}$Department of Applied Physics and Geballe Laboratory for Advanced Materials, Stanford University, California 94305, USA\\
$^{4}$Stanford Institute for Materials and Energy Sciences, SLAC National Accelerator Laboratory, 2575 Sand Hill Road, Menlo Park, California 94025, USA}
\date{\today}

\begin{abstract}
We present detailed calculations of the electric field gradient (EFG)  using a point charge approximation  in  \BaOs, 
a  Mott insulator with strong spin-orbit interaction.    Recent \Na nuclear magnetic resonance (NMR) measurements found that the onset of  local point symmetry breaking, likely caused by the formation of quadrupolar order \cite{ChenBalents10}, precedes  the formation of long range magnetic order in this compound \cite{Lu17, Liu17}. 
 An extension of the  static \Na NMR measurements as a function of the orientation of  a 15 T applied magnetic field at 8 K in the magnetically ordered phase is reported.  
 Broken local cubic symmetry induces a non-spherical electronic charge distribution around  the Na site and thus finite EFG, affecting the NMR spectral shape.  
 We combine the spectral analysis as a function of the orientation of the magnetic field with calculations of the EFG  to determine the exact microscopic nature of the lattice distortions present in low temperature phases of  this material. We establish that orthorhombic distortions, constrained along the cubic axes of the perovskite reference unit cell,  of oxygen octahedra  surrounding Na nuclei  are present in the magnetic phase.  
 Other common types of distortions often observed in oxide structures are considered as well.

 %\item[DOI]    
 
\end{abstract}

\maketitle

\section{\label{intro}INTRODUCTION}

The investigation of the effects of spin orbit coupling (SOC) is one of the central issues in  the study of quantum materials \cite{KrempaRev14}. In addition to its key role in inducing topological phases,  the  combined effects  of SOC and strong electronic correlations can lead to numerous emergent quantum phases \cite{Jackeli09, ChenBalents10,Chen11,Dodds11,Cole12, KrempaRev14, Svoboda17,Balents17}.  A theoretical description of these phases   is challenging. 
Certain  approaches based on   multipolar interactions have been proposed \cite{ChenBalents10, Balents14,Svoboda17, Balents17}. 
The key prediction of the quantum models with multipolar magnetic interactions is that %a %quadrupolar/orbitally ordered phase precedes the formation of long range magnetism. 
a structural symmetry is lowered in the magnetically ordered phase. In fact for specific parameters,   a quadrupolar/orbitally ordered phase precedes the formation of long range magnetism \cite{ChenBalents10, Balents14,Svoboda17}. 
To provide tests of such theory one needs a probe that is concurrently sensitive to both orbital and spin degrees of freedom. %This is to assure that symmetry breaking  prefaces  the magnetism regardless of the span of the  temperature range in which the orbital order precedes magnetism. 
Nuclear magnetic resonance (NMR)   on nuclei with asymmetric charge distributions provide such tests, as was shown in \cite{Lu17, Liu17}.
 In fact, in our recent \Na NMR measurements of  the Mott insulator with strong spin-orbit interaction \BaOs,   we reported that  the onset of  local point symmetry breaking, likely caused by the formation of quadrupolar order, precedes  the formation of long range magnetic order \cite{Lu17, Liu17}.

Here, we present  an analysis of the angular evolution of NMR data as an applied magnetic  field is rotated in different plains of the crystal. This analysis led us to conclude that the  symmetry lowering transition is to an orthorhombic  point symmetry.  In crystals with cubic symmetry, the electric field gradient (EFG) vanishes. The lowering of the symmetry induces  a finite EFG, which is a quantity directly observable in a static  NMR measurement on a nuclei with  finite  quadrupole moment, such as \NaE. Specifically, the parameter extracted from the spectra of such NMR  experiments is the quadrupole resonance frequency, 
defined in terms of
  $V_{zz}$,  which is  the largest principal component of the EFG at the nuclear site, and other  intrinsic nuclear properties \cite{Cohen57, Kaufmann79,Alex2004}.  
Since the EFG tensor is a traceless rank-two tensor, its components can be determined by analysis of the spectra obtained as the orientation of the magnetic field is rotated with respect to the crystalline axes \cite{Cohen57,   Itoh03}. 
%\cite{Tokunaga06, BerthierCDW}

We use EFG calculations based on a point charge approximation \cite{Cohen57} to describe how various local lattice deformations affect \Na spectra in \BaOs. 
A comparison of these results with experimental findings allows us to determine the microscopic nature of local cubic symmetry breaking. In particular, we determine that  the  broken symmetry phase is characterized by the distortions of oxygen octahedra involving 
 dominant displacement of oxygen ions along the cubic axes of the perovskite reference unit cell.  Our work illustrates  that,  in addition to being a well known powerful probe of local magnetism and/or charge density waves \cite{BerthierCDW},  NMR is a good probe of spatial point symmetry breaking as well.

% In  NMR experiments on nuclei with finite  quadrupole moment, such as \Na, the %parameter extracted from the spectra is the quadrupole resonance frequency, defined in %terms of
%where $V_{zz}$,  the largest principal component of the EFG at the nuclear site and other %intrinsic nuclear properties. 

The remainder of the paper is organized as follows. In Sec. II, we give a basic overview of the quadrupole interactions and   present the ways in which these quadrupole effects  are manifested in \BaOs. We present the angle dependence of the quadrupole splitting, \ie frequency difference between two adjacent quadrupole perturbed Zeeman energy levels,    
as applied magnetic field is rotated in two different planes of the crystal  in Sec. III. 
In Sec. IIIA  we present 
detailed analysis of the angular dependence data used to determine the exact symmetry    of the EFG.  The  point charge approximation approach for calculating EFG is introduced  in Sec. IV.  Results of the point charge calculations for various distortion models are presented in  Sec. V.

\section{\label{Sec1}Manifestation of Quadrupole Effect in B{\MakeLowercase a}$_2$N{\MakeLowercase a}O{\MakeLowercase s}O$_6$ }

The quadrupole effect   refers to  the interaction between the non-spherical nuclear charge distribution and an electrostatic field external to the nucleus.  The non-spherical nuclear charge distribution appears in nuclei with spin $I > 1/2$ and is represented by the nuclear quadrupole moment $\hat Q$ operator, a second-rank tensor defined by the integral over the nuclear charge distribution \cite{Cohen57}. This operator can be more conveniently expressed in terms of the the nuclear spin operators $I$. In this case, its magnitude is proportional  to what is conventionally referred to as the nuclear quadrupole moment $eQ$.     
The relevant electrostatic field, assuming a Laplacian potential, is represented by the  EFG generated at the nuclear site by surrounding electronic charges  \cite{Cohen57, Kaufmann79}. Therefore, the strength of the quadrupolar interaction is dictated by the product of the  nuclear quadrupole moment  and the magnitude of the EFG. The nuclear quadrupole moment   is non-zero for nuclei with spin  $I > 1/2$, while the EFG is non-vanishing for point charges arranged on a lattice with symmetry lower than cubic. Thus, quadrupolar interactions generates finite effects only if $I>1/2$ and  the electronic charge distribution is asymmetric (non-cubic).

\begin{center}	
 %
%%%%%%%%%%%%%%%%%%%%%%%%%%%%%%%%%%%%%%
\begin{figure}[t]
  \vspace*{-0.0cm}
\begin{minipage}{0.98\hsize}
%%%%%%%%%%%%%%%%%%% F I G U R E 1 %%%%%%%%%%%%%%%%%%%%
 \centerline{\includegraphics[scale=0.41]{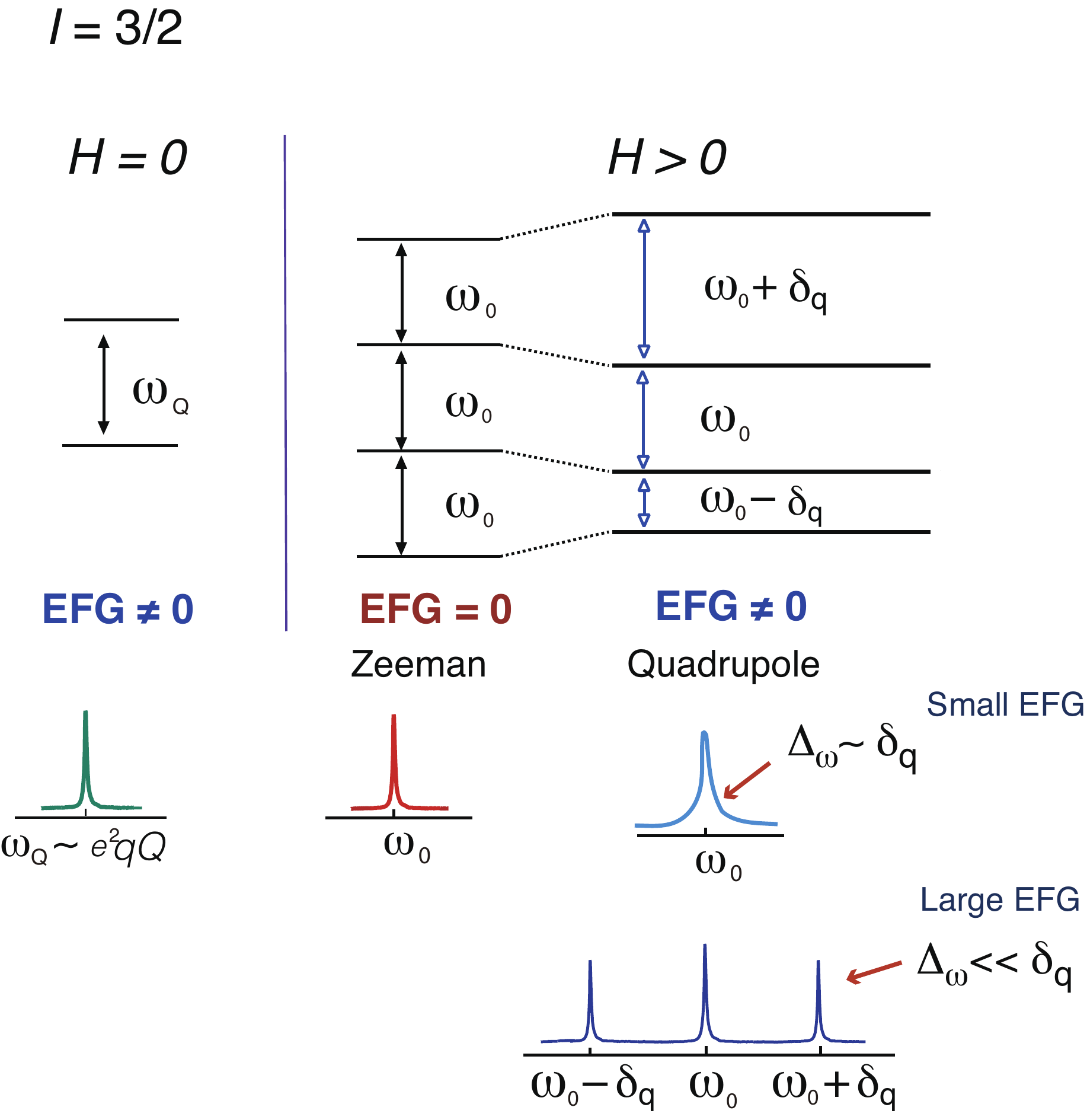}} %%%%%%%%%%%%%%%
%%%%%%%%%%%%%%%%%%%%%%%%%%%%%%%%
\begin{minipage}{.98\hsize}
 \vspace*{-0.0cm}
\caption[]{\label{FigLev} \small %(Color online)  
Schematic of the energy levels for a $I$=3/2 nucleus in both zero and a finite magnetic field $H$,    in the presence of quadrupole interaction with the EFG  generated by surrounding electronic charges, and the resulting NMR spectra. In principle, zero applied field and a finite EFG  result in  a single  line at frequency $\omega_{Q}$ that can be observed in a  nuclear quadrupole resonance (NQR) experiment. The frequency   $\omega_{Q}$ is proportional to the product of nuclear quadrupole moment and the magnitude of the EFG. 
In a finite applied field and in the absence of quadrupole interaction, the  spectrum consists of a single narrow line at frequency $\omega_{0}$.  
In the presence of a quadrupole interaction  that acts as a perturbation to dominant Zeeman Hamiltonian in the depicted case,  the central transition remains at frequency  $\omega_{0}$.   The satellite transitions appear at frequencies    shifted  by $\pm$ \dq, which proportional to the magnitude of the EFG (\mbox{Eq. \ref{delGen}}). For small values of the EFG, satellite transition cannot be resolved and only line broadening is observed. 
Strictly speaking, there is also a broadening due to the distribution of magnitude of the EFG  itself, but this is manifested only on the satellites and not on the central transition. %In our case, this distribution can be neglected as all the lines show the same width.
}
 \vspace*{-0.3cm}
\end{minipage}
\end{minipage}
\end{figure}
%%%%%%%%%%%%%%%%%%%%%%%%%%%%%%%%%%%%%%
%
\end{center}

In \mbox{Fig. \ref{FigLev}}  we show a schematic of the energy levels for nuclei with spin  $I$=3/2, such as \Na that we investigated. Energy levels are displayed     in both zero and a finite magnetic field $H$ and in the presence of quadrupole interaction with the EFG. The   resulting NMR spectra are also shown. As it was the case in our experiments, in a finite field we represent the quadrupolar interaction as a perturbation to the dominant Zeeman term. In zero applied field and in the presence of a finite EFG,  a single  line   at  a frequency  proportional to the product of the nuclear quadrupole moment and the magnitude of the EFG   can be observed in a  nuclear quadrupole resonance (NQR) experiment. In a finite applied field and in the absence of quadrupolar interaction (\ie EFG=0), the spectrum consists of a single narrow line at the resonant frequency $\omega_{0}$, which is proportional to the magnitude of the applied field.   In the presence of quadrupolar interaction (\ie \mbox{EFG $\neq 0$}) the central transition remains at frequency  $\omega_{0}$. The satellite transitions appear at frequencies    shifted  by $\pm$ \dq, which are proportional to the magnitude of the EFG (\mbox{App. \ref{QuadApp}}). 
Thus, quadrupole interaction splits otherwise single NMR line to 2$I = 3$ lines. 
Satellite transition cannot be resolved and only line broadening is observed 
for small values of the EFG. 
  
  \vspace*{-0.90cm}

\begin{center}	
 %
%%%%%%%%%%%%%%%%%%%%%%%%%%%%%%%%%%%%%%
\begin{figure}[t]
  \vspace*{-0.0cm}
\begin{minipage}{0.98\hsize}
%%%%%%%%%%%%%%%%%%% F I G U R E 2 %%%%%%%%%%%%%%%%%%%%
 \centerline{\includegraphics[scale=0.69]{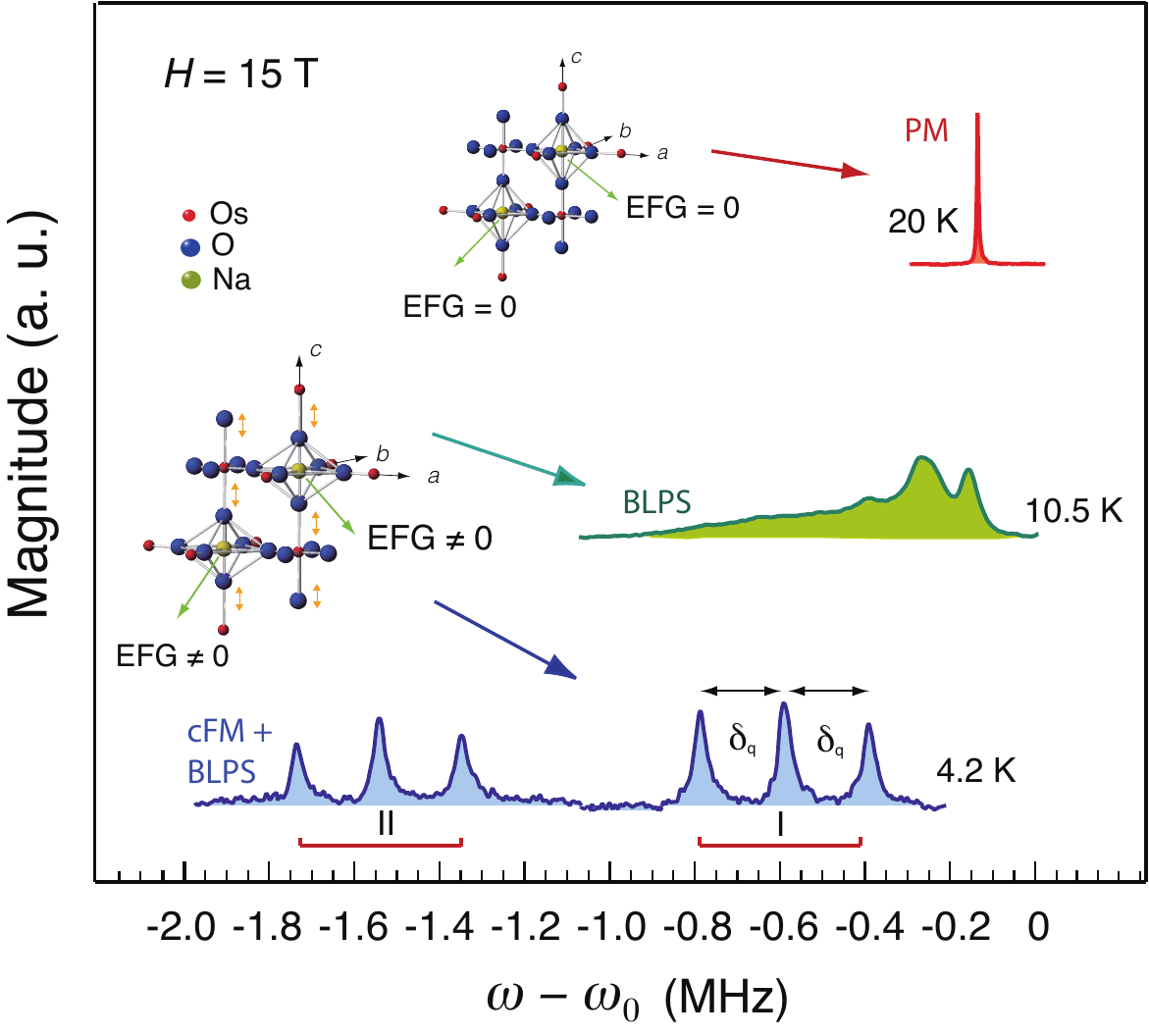}} %%%%%%%%%%%%%%%
%%%%%%%%%%%%%%%%%%%%%%%%%%%%%%%%
\begin{minipage}{.98\hsize}
 \vspace*{-0.2cm}
\caption[]{\label{FigSpec} \small %(Color online)  
 Temperature evolution of $^{23}$Na in Ba$_2$NaOsO$_6$ spectra at 15 T  with magnetic field applied parallel to [001] crystallographic axis. %Data is adopted from \cite{Liu17}. 
 At 20 K, 
 a narrow single peak spectrum characterizes the  high temperature paramagnetic (PM) state. In this state,  the crystal structure of \BaOsS is undistorted, as depicted. That is, point symmetry at the Na site is cubic and leads to a  to vanishing EFG.  
 At intermediate temperatures, broader and more complex spectra reveal the appearance of finite EFG  induced by the breaking  of local cubic symmetry. In this case, the crystal structure of \BaOsS is distorted so that point symmetry at the Na site is non-cubic, inducing a finite  EFG.  
At lower temperature, the splitting into 2 sets of triplet lines (labeled as I and II)   reflects the existence of two distinct magnetic sites in the lattice.
 Zero of frequency is defined as $\omega_{0} =  \, ^{23}\gamma \,H$.  Splitting between quadrupolar satellites is denoted by $\delta_{q}$. Abbreviation: PM, paramagnetic; BLPS, broken local  point symmetry; and, cFM,  canted ferromagnetic. 
}
 \vspace*{-0.3cm}
\end{minipage}
\end{minipage}
\end{figure}
%%%%%%%%%%%%%%%%%%%%%%%%%%%%%%%%%%%%%%
%
\end{center}

 These schematic qualitative  describe our \Na NMR observations in \BaOs.  
 The main effects of the quadrupole interaction observed in this compound, are illustrated in \mbox{Fig. \ref{FigSpec}}. 
 The high temperature, 20 K,   paramagnetic state spectra consist of a single narrow NMR line. Since   the nuclear spin for $^{23}$Na equals to $3/2$, the absence of the three distinct quadrupolar satellite lines indicates that the EFG is zero as a consequence of  a  cubic environment.  
Lowering the temperature broadens  the NMR line (e.g. at 10.5 K) and eventually splits it  into multiple peaks (e.g. at 4.2 K). This splitting  indicating the start   of significant changes in  
 the local symmetry, thereby producing an EFG, \ie asymmetric (non-cubic) charge distribution. 
 Therefore, the observed line broadening and subsequent splitting of the Na spectra into triplets  in the magnetically    ordered phase  indicates breaking of the  cubic point symmetry  caused by local distortions of electronic charge distribution, as established in \cite{Lu17, Liu17}. 
 The data was taken in the same experimental conditions as described in detail in \mbox{Ref. \cite{Liu17}}. 
  At   4.2 K, the  \Na spectra clearly split into 6 peaks.  These peaks correspond to   two sets of triplet lines,   labeled  as I and II in \mbox{Fig. \ref{FigSpec}}, that are well separated in frequency.  
 As previously established,  these two sets of  triplets   appear due to  magnetic interactions \cite{Lu17}  that are irrelevant to our discussion of quadrupole effects.   The splitting labeled   \dqS in \mbox{Fig. \ref{FigSpec}}   implies that a finite EFG has been  induced by changes in local charge distribution. In this paper we will consider various modifications of  local lattice symmetry that can induce a  finite EFG and account for our experimental observations \cite{Lu17, Liu17}.   
 However, we first give a more quantitative overview of the quadrupole interaction.

  \vspace*{-0.10cm}

 For anisotropic charge distributions, the quadrupole Hamiltonian expressed in the coordinate system define by the principal axes of the EFG  is given by
\begin{equation}
\begin{split}
\label{HamQ1}
\mathcal{H}_{\rm Q}(x,y) = \frac{eQV_{zz}}{4I(2I-1)} \left[(3\hat{I}^2_{z}-\hat{I}^2)+\eta(\hat{I}^2_{x}-\hat{I}^2_{y}) \right ],
\end{split}
\end{equation}
where   $eQ$ is the quantity conventionally referred to as the nuclear quadrupole moment, $\eta \equiv\left |  {V_{xx}-V_{yy}} \right |/{V_{zz}}$  is the asymmetry parameter,  and  V$_{\rm xx}$, V$_{\rm yy}$, and V$_{\rm zz}$ are diagonal components of the EFG.
Here,  $V_{\rm zz}\equiv eq$ is defined as the principle component of the EFG and  $|V_{\rm xx}|< |V_{\rm yy}|< |V_{\rm zz}|$, by convention \cite{AbragamBook}. 
The EFG   is a symmetric and traceless $3 \times 3$ tensor that corresponds to the rate of change of the electric field at an atomic nucleus  \cite{Kaufmann1979}.    The principal axis of the EFG define 
   the coordinate system $O_{XYZ}$, which is not necessarily aligned with that defined by the crystalline axes $O_{xyz}$. Evidently, $V_{zz}$ is parallel to one of the crystal axes if 
the principal axes of the EFG and those of the crystal are aligned.

We define the observable  \dq to  represent the quadrupole splitting  between different quadrupole satellites. As derived in \mbox{App.\ref{QuadApp}}, $\delta_{\rm q}$ corresponds to the  frequency difference between adjacent  quadrupole satellite transitions.  
In the most general case the quadrupole splitting $\delta_{\rm q}$ is given by
\begin{equation}
\label{delGen}
\delta_{\rm q} = \frac{(eQ)( V_{\rm zz})}{2 h}   \, \left ( 1 +   \frac{\eta^{2}}{3} \right )^{{1/2}}.  
\end{equation}
Thus, the value of  $\delta_{\rm q}$ is dictated by both  the magnitude of the principal component of the EFG $(V_{\rm zz})$ and the anisotropy parameter $\eta$. In the high field limit, when $\mathcal{H}_{\rm Q}$ is a perturbation to the dominant Zeeman term, the angular dependence of the splitting is given by
\begin{equation}
\delta_{\rm q} = \frac{\nu_{\rm q}}{2} \left ( 3\cos^2\theta - 1 + \eta \sin^2\theta \cos2\phi \right),
\label{qsplit}
\end{equation}
where   $\theta$ is the angle  between the applied field $ H$ and $V_{zz}$,  $\phi$ is the standard azimuthal angle of a spherical coordinate system defined by $O_{{XYZ}}$, and  $\nu_Q \equiv   \frac{(eQ)V_{{zz}}}{2h}$.    
 Therefore, to test whether an observed splitting in the   NMR spectra originates from quadrupole effects, one has to measure the spectra as  a function of strength and orientation of the applied magnetic field.  
 Clearly, for a fix orientation of the applied field  the splitting should be independent of its magnitude.  As a matter of fact, we establish that \dqS varies by $\le 2\%$, which is if the order of the error bars, as $H$ increases from 7 T to 29 T \, \cite{Liu17}.  Insensitivity of \dqS   to the strength of the magnetic field implies that the splitting originates from quadrupole effects. Namely,   the  finite EFG is induced by changes in charge density distribution and/or lattice distortions and not by trivial magnetostriction effect on a crystal.  However, to decipher  the detailed structure of the    EFG tensor, one has to investigate how \dqS evolves as the  orientation of the applied field is varied with respect 
 to crystalline axes.  As we will describe in detail in the next section, this type of rotational studies allow us to discern the exact nature of the distortions.

 \section{ANGULAR DEPENDENCE OF QUADRUPOLE EFFECT}

 \begin{center}	
 %
%%%%%%%%%%%%%%%%%%%%%%%%%%%%%%%%%%%%%%
\begin{figure}[t]
  \vspace*{-0.0cm}
\begin{minipage}{0.98\hsize}
%%%%%%%%%%%%%%%%%%% F I G U R E 3 %%%%%%%%%%%%%%%%%%%%
 \centerline{\includegraphics[scale=0.73]{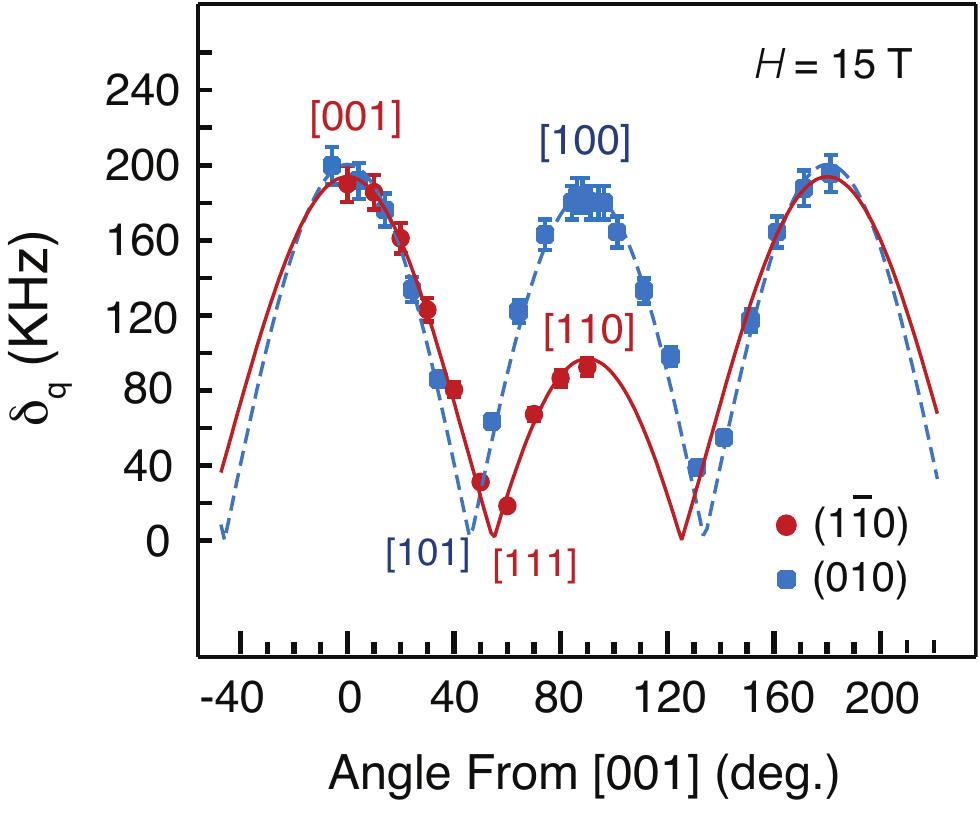}} %%%%%%%%%%%%%%%
%%%%%%%%%%%%%%%%%%%%%%%%%%%%%%%%
\begin{minipage}{.98\hsize}
 \vspace*{-0.0cm}
\caption[]{\label{FigRotSp} \small %(Color online)  
    The mean peak-to-peak splitting $(\delta_{q})$ between any two adjacent peaks of the triplets I and II as a function 
of the angle between [001] crystal axis and the applied magnetic field $(H)$.  
 The red circles denote angular dependence of splitting for $H$ rotated  in $(1 \bar1 0)$ plane. The red solid line is the fit to $|(3   \cos^{2} \theta -1)/2 |$, where $\theta$ denotes the angle between the principal axis of the EFG $(V_{ZZ})$ and the applied magnetic field.  The blue squares denote the angular dependence of splitting when sample is rotated  in the $(010)$ plane. The blue dotted line is the fit to $|(3   \sin^{2} \theta -1-\eta\cos^{2} \theta)/2 |$ \mbox{(Eq. \ref{dq4})}, where $\theta$ denotes the angle between the  $(V_{ZZ})$ and $H$, and $\eta$ is the sasymmetric parameter as explained in the text.
}
 \vspace*{-0.3cm}
\end{minipage}
\end{minipage}
\end{figure}
%%%%%%%%%%%%%%%%%%%%%%%%%%%%%%%%%%%%%%
%
\end{center}
  \vspace*{-0.90cm}

We   performed detailed measurements of \dqS at 15 T and 8 K as a function of the angle $(\theta)$ between $H$ and the [001] crystalline axis   in two different planes of the crystal ($(1 \bar 1 1)$ and $(010)$). The measured angle dependence of the splitting is plotted in \mbox{Fig. \ref{FigRotSp}}.   We observe that the splitting between any two adjacent peaks of the triplets I and/or II is equal, within the error bars. Therefore,  we plot the mean peak-to-peak splitting    between any two adjacent peaks of the triplets I and II. 
We observed  that the splitting is the largest for $H \| [001]$. Moreover, for rotations in the $(010)$ plane, \ie along one face of the cubic unit cell, we find that \dqS reaches its maximum value for    $H \| [100]$  as well. 
 Furthermore as   described in  \mbox{Ref. \cite{Lu17}}, we observe no more than 3 lines  per set (I or II) regardless of the angle $\theta$.  This indicates that the magnetic field was rotated in the coordinate system defined by the principal axes of the EFG. In other words, the principal axes of the EFG must coincide with those of the crystal in a low temperature non-cubic phase of  \BaOs. This observation  together with finding equal \dqS on triplets I and II, \ie two magnetically inequivalent Na sites, implies that in the simplest scenario the finite EFG arises from distortions of the O$^{2-}$ octahedra surrounding Na$^{+}$ ions with the oxygen constrained to move along the cubic axes of the perovskite reference unit cell, as illustrated in \mbox{Fig. \ref{FigSpec}}.

In a material with global cubic symmetry  such as \BaOs,  it is thus possible to stabilize three different domains, each with the   principle axis of the EFG, $V_{zz}$, pointing along any of the three equivalent crystal axes.  %Furthermore, local magnetic field has to be parallel to $V_{\rm zz}$ in each domain. 
 The fact that the splitting is the largest for $H \| [001]$, 
 and that only three peaks per set are observed for $H \| [110]$ imply that two domains are plausible   in the crystal.  One domain  is characterized by  pure uniaxial $3z^2 - r^2$ distortions where $V_{\rm zz}$   is in [001] direction, while   the other is distinguished by $x^2 - y^2$ distortions where $V_{\rm zz}$ is  in the (110) plane.  
In the simplest case $V_{\rm zz}$ is parallel  to the [001] direction with $\eta = 0$, indicating tetragonal local symmetry. In the second case, $V_{\rm zz}$ is aligned along the [100] direction with  $\eta$  of  order  1,   implying orthorhombic local symmetry.  
To determine the exact local symmetry, \ie distinguish between tetragonal  and orthorhombic distortions, we need to consider the details of the angular dependence of  the splittings $\delta_{\rm q}$ obtained  
for rotations of the applied field in the $(010)$ plane. The fact  that \dqS reaches its maximum value for    both $H \| [001]$ and $H \| [100]$,   for the field rotated in the  the $(010)$ plane,   reveals orthorhombic local symmetry.
Next, we discuss in detail the claim that our observations imply orthorhombic local symmetry.

\subsection{Tetragonal Symmetry vs. Orthorhombic Symmetry}

To qualitatively analyze \dqS as the magnetic field is rotated, we start by transforming the  Hamiltonian in \mbox{Eq. \ref{HamQ1}} into the coordinate system $O_{{\rm xyz}}$, defined by  $H$ having the  $O_{\rm z}$ axis parallel to the applied field.  Assuming that the local symmetry is tetragonal, that is $V_{xx} = V_{yy}$ and $\eta = 0$, and that $V_{zz}$ is parallel to crystalline $c$ axis, we may without loss of generality choose the axis, $O_z$ parallel to the applied magnetic field $H$ in plane XOZ, so that 
\begin{equation}
I_Z = I_z\cos\theta + I_x\sin\theta,
\end{equation}
where $\theta$ is the angle between $H$ and $V_{zz}$. 
Then, the quadrupole Hamiltonian in O${_{\rm xyz}}$ becomes  
\begin{equation}
\begin{split}
\mathcal{H}_Q =& \frac{1}{6}h\nu_Q \Large\{ \frac{1}{2}(3\cos^2\theta-1)(3I^2_z-I(I+1))\\&+ \frac{3}{2}\sin\theta \cos\theta[I_z(I_++I_-) + (I_++I_-)I_z] \\&+ \frac{3}{4}\sin^2\theta(I^2_++I^2_-)   \}, 
\label{eq:quadrupoleHamiltonian}
\end{split}
\end{equation}
where  $\nu_Q = \frac{3e^2qQ}{2hI(2I-1)}$.
\begin{center}	
 %
%%%%%%%%%%%%%%%%%%%%%%%%%%%%%%%%%%%%%%
\begin{figure}[h]
  \vspace*{-0.0cm}
\begin{minipage}{0.98\hsize}
%%%%%%%%%%%%%%%%%%% F I G U R E 4 %%%%%%%%%%%%%%%%%%%%
 \centerline{\includegraphics[scale=0.67]{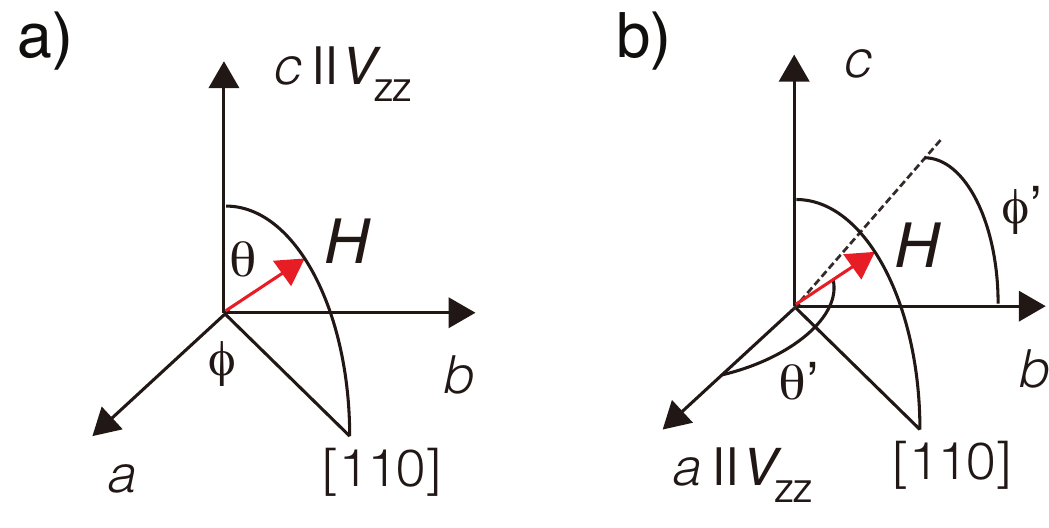}} %%%%%%%%%%%%%%%
%%%%%%%%%%%%%%%%%%%%%%%%%%%%%%%%
\begin{minipage}{.98\hsize}
 \vspace*{-0.0cm}
\caption[]{\label{FigRot} \small %(Color online)  
Schematic of rotation of the applied field in $(1 \bar1 0)$ plane. The red arrow denotes the applied field direction. 
 $V_{ZZ}$ is parallel to $c$  and $a$  (or $b$ ) axis in $\bf a)$ and  $\bf b)$, respectively. 
  }
 \vspace*{-0.3cm}
\end{minipage}
\end{minipage}
\end{figure}
%%%%%%%%%%%%%%%%%%%%%%%%%%%%%%%%%%%%%%
%
\end{center}
  \vspace*{-0.20cm}
 Taking the quadrupole Hamiltonian as a perturbation to dominant Zeeman term, the energy eigenstates of $\mathcal{H}_Q$ are given by
 \begin{equation}
 E_{m} =   \frac{1}{12}h\nu_Q\, [3m^2-I(I+1)](3\cos^2\theta-1).
 \end{equation}
The quadrupole splitting $\delta_{q}$ between adjacent quadrupole satellites then equals to 
\begin{equation}
\begin{split}
\delta_{q} = \frac{ E_{m} -   E_{m-1}}{h} =  \frac{1}{2}h\nu_Q \,(3\cos^2\theta-1).   
\end{split}
\label{dq1}
\end{equation}
  \mbox{Eq. \ref{dq1}} describes the angular dependence of quadrupole splitting in the case of tetragonal symmetry ($\eta = 0$) and $V_{zz}$ parallel to c-axis. We observe such angular dependence 
when $H$ is rotated from the [001] to [110] direction, \ie in the $(1 \bar 1 1)$ plane, as shown in 
in \mbox{Fig.\ref{FigRotSp}}.

If $\eta$ is not confined to be zero, one deduces a more general form of  the energy eigenstates of $\mathcal{H}_Q$ \cite{AbragamBook}, 
\begin{equation}
\begin{split}
E_m = \frac{1}{12}h\nu_Q & [3m^2 - I(I+1)] \times \\ 
 &[(3\cos^2\theta - 1) +\eta \sin^2\theta \cos2\phi ], 
\end{split}
\end{equation}
leading to 
\begin{equation}
\begin{split}
\delta_q &= \frac{E_{m} - E_{m-1}}{h}    \\
& = \frac{1}{2}\nu_Q \left(3\cos^2\theta - 1 + \eta \sin^2\theta \cos2\phi  \right), 
\label{dq2}
\end{split}
\end{equation}
where angles $\theta$ and $\phi$ are as defined in \mbox{Fig. \ref{FigRot}}. 
For the applied field  rotated in the $(1 \bar1 0)$ plane, as was the case in one of our measurements,  $\cos2\phi$ = 1, as illustrated in  \mbox{Fig. \ref{FigRot}a)}. Then, fitting our data for angular dependence of \dqS as $H$ is rotated in the $(1 \bar1 0)$ plane to  \mbox{Eq. \ref{dq2}},  we again obtain that  $\eta = 0$. 

Up to this point, we have assumed that $V_{zz}$ is parallel to the $c$ axis. 
We point out that the splitting \dq, as derived in all of the above equations, depends on polar angles  $\theta$ and $\phi$ that are given in the coordinate system defined by the principal axes of the EFG. 
When $V_{zz}$ is parallel to $c$ axis,   the coordinate system   defined by the principal axes of the EFG coincides with  that defined by crystalline axes.   However, if $V_{zz}$ is parallel to  the $a$ or $b$ axis as depicted in \mbox{Fig. \ref{FigRot}b)}, $\theta$ and $\phi$  need to be transformed into crystalline coordinates.  
This transformation is necessary for a meaningful comparison with the data as only the orientation of $H$ with respect to crystalline axes is know in an experiment. 
We denote 
angles $\theta'$ and $\phi'$ (which are $\theta$ and $\phi$ in \mbox{Eq. \ref{dq2}}) as angles defined in the 
 EFG coordinate system.    These angles are related to the angles  $\theta$ and $\phi$   in the crystalline coordinate system according to the following transformations, 
\begin{equation}
\begin{split}
&\cos\theta' = \sin\theta \cos\phi  \\
&\sin\theta' \cos\phi'= \cos\theta\\
&\sin\theta'\sin\phi'= \sin\theta \sin\phi .
\end{split}
\end{equation}
When the  applied  magnetic field is rotated in the (1$\bar{1}$0) plane $\phi$ = 45$^\circ$, as shown in \mbox{Fig. \ref{FigRot}b)},    \mbox{Eq. \ref{dq2}} becomes  
 \begin{equation}
\delta_q = {1 \over 2} \nu_Q \left ( \frac{3}{2}\cos^2\theta - 1 + \eta \left ( \frac{3}{2}\cos^2\theta-\frac{1}{2}  \right)\right). 
\label{rot_angle}
\end{equation}
Evidently, fitting the angular dependence of \dqS for $H$ rotated  in the (1$\bar{1}$0) plane to  \mbox{Eq. \ref{rot_angle}}  produces  the same quality fitting curve as a fit to  \mbox{Eq. \ref{dq2}}   but with $\eta \approx 0.87$. The non-zero value of  $\eta$  indicates that the symmetry is lower than tetragonal. Moreover, as  $V_{zz}$ is parallel to either $a$ or $b$-axis,   symmetry must be orthorhombic.

\begin{center}	
 %
%%%%%%%%%%%%%%%%%%%%%%%%%%%%%%%%%%%%%%
\begin{figure}[t]
  \vspace*{-0.0cm}
\begin{minipage}{0.98\hsize}
%%%%%%%%%%%%%%%%%%% F I G U R E 5 %%%%%%%%%%%%%%%%%%%%
 \centerline{\includegraphics[scale=0.33]{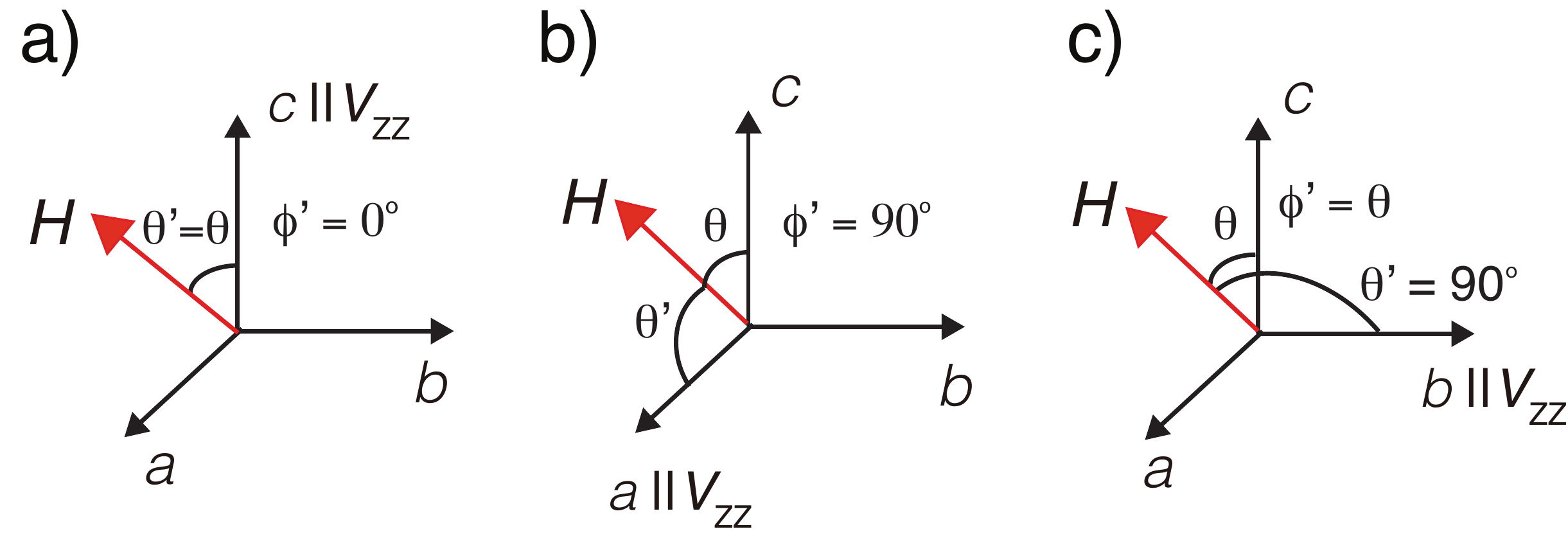}} %%%%%%%%%%%%%%%
%%%%%%%%%%%%%%%%%%%%%%%%%%%%%%%%
\begin{minipage}{.98\hsize}
 \vspace*{-0.0cm}
\caption[]{\label{FigRot2} \small %(Color online)  
 Schematic of rotation of the applied field in $(010)$  plane. The red arrow denotes the applied field direction. 
 $V_{ZZ}$ is parallel to $c$,  $a$, and $b$ axis in $\bf a)$, $\bf b)$, and  $\bf c)$, respectively. 
}
 \vspace*{-0.3cm}
\end{minipage}
\end{minipage}
\end{figure}
%%%%%%%%%%%%%%%%%%%%%%%%%%%%%%%%%%%%%%
%
\end{center}

Thus far, both the tetragonal symmetry (with either $V_{zz}$   parallel to $c$ axis or $a(b)$ axis) and orthorhombic symmetry (with $V_{zz}$  parallel to $a$ or $b$ axis) EFG could account for the observed angular dependence of \dqS when $H$ is rotated  in the $(1 \bar1 0)$ plane. Clearly, in orthorhombic structure, C4 rotation symmetry is broken and the oxygen  octahedra are distorted so that $a \neq b \neq c$. To  determine  undeniably whether distortions (\ie EFG) are tetragonal or orthorhombic, we performed another  measurement of the angular dependence of the splitting in which the applied field was rotated in the (010) plane. To understand   these results,     plotted in  \mbox{Fig. \ref{FigRotSp}},   we consider different orientations of $V_{zz}$ as shown in \mbox{Fig. \ref{FigRot2}}.
  \vspace*{-0.90cm}
 First, assuming that $V_{zz}$ is parallel to  the $c$ axis,   $\theta' = \theta$ and $\phi' = 0$,  so that \mbox{Eq. \ref{dq2}}    becomes, 
\begin{equation}
\delta_q = \frac{1}{2}\nu_Q(3\cos^2\theta - 1 + \eta \sin^2\theta). 
\label{dq3}
\end{equation}
If on the other hand $V_{zz}$ is parallel to the $a$ axis,  we obtain    $\theta' = 90^\circ - \theta$ and $\phi' = 90^\circ$. The angular dependence of splitting is given by,
\begin{equation}
\delta_q = \frac{1}{2}\nu_Q(3\sin^2\theta - 1 - \eta \cos^2\theta).
\label{dq4}
\end{equation}
Finally, if  $V_{zz}$ is parallel to the $b$ axis, we have $\theta' = 90^\circ$ and $\phi' = 0$. The splitting   becomes,
\begin{equation}
\delta_q = \frac{1}{2}\nu_Q(\eta \cos2\theta - 1).
\label{dq5}
\end{equation}
The observed angular dependence of \dqS for $H$ rotated  in the (010) plane can be fit to both 
\mbox{Eq. \ref{dq3} \& \ref{dq4}}. In both cases,  the fitting yields $\eta \approx 0.87$.

To sum up, the observed angular dependence of the splitting confirms that such splitting is due to quadrupole effect. Since no more than 3 lines  per set (I or II) are observed regardless of the angle $\theta$, the principal axes of the EFG must coincide with those of the crystal. Furthermore, we found that distortions with both tetragonal   (with $V_{zz}$  parallel to $c$-axis or $a(b)$-axis) and orthorhombic symmetry (with $V_{zz}$   parallel to either $a$ or $b$-axis) could account for angular dependence of the splitting for $H$ rotated in the  $(1 \bar1 0)$ plane.  However, only orthorhombic distortions (with $V_{zz}$  parallel to $c$-axis or $a$-axis) could explain  the  angular dependence of the splitting \dqS for $H$ rotated in the $(010)$ plane. Therefore, by combining  
the results of the angular dependence of \dqS  for  $H$ rotated  in   two different planes, we conclude that  
the  distortion is orthorhombic with $\eta \approx 0.87$ and $V_{zz} \|  a$.

 As we described,  in a material with global cubic symmetry,  it is   possible to stabilize three different domains, each with the   principle axis of the EFG, $V_{zz}$, pointing along any of the three equivalent crystal axes. 
Therefore, for either a cubic to orthorhombic, or a tetragonal to orthorhombic phase transition, formation of distinct  domains, with their principle axes rotated by 90 degrees, is expected. The analysis of the angular dependence of the spectral lineshapes in  
the low temperature magnetic phase     did not provide any evidence for the formation of different domains. Thus, this must imply    the presence of some weak symmetry-breaking field that favors one domain over the others. We believe that this symmetry-breaking field  is provided by the strain from the way the sample was mounted on the flat platform, which was always parallel to the specific face of the crystal.

\section{Point Charge Approximation}

 We employed the point charge model to calculate $V_{ZZ}$ and  $\eta$ resulting from different distortion scenarios. This is done to find the full set of possible distortions that can account for our observations, \ie maximum splitting equals $\delta_{q} \approx 190 \, \rm{kHz}$, for $H \| c$  for all satellite transitions, $\eta \approx 0.87$, and $V_{ZZ} \|   a$.
 %
%
%To find the full set of possible distortions that can account for our observations that  maximum splitting equals $\delta_{q} \approx 190 \, \rm{kHz}$, for $H \| c$  for all satellite transitions, $\eta \approx 0.87$, and $V_{ZZ} \|   a$, we used the point charge model to calculate $V_{zz}$ and $\eta$ resulting from different distortion scenarios. 
In this model, the electron density at the Na site is calculated by taking into account all the surrounding charges,  which are treated as the point charges of zero radius that carry the appropriate ionic charge.  
The surrounding charges are accounted for by a lattice summation method that is easily employed for systems with large number of atoms and/or single crystals. 
We note that the point-charge approximation of ions neglects any covalent nature of the bonding in a material and is therefore strictly valid in strongly ionic compounds, which is the case in \BaOs. 
The EFG tensor components $V_{ij}$,  $i,j=x,y,z$,  at a certain nuclear site resulting from an ion of charge  $q$    are given  by \cite{Herzig1985}
\begin{equation}
\label{EFG_cal}
V_{ij} = \sum_\mu{\sum_k{q_{k}\frac{3({}^{k}r_{i}^{\mu})({}^{k}r_{j}^{\mu})-\delta_{ij}({}^{k}r^{\mu})^2}{({}^{k}r^{\mu})^5}}} , 
\end{equation}
where $\sum_\mu$ denotes sum over multiple unit cells, and $\sum_k$ is the sum  over all atoms within a single unit cell,  and $r$ is   the distance from the specific nuclear site, the point of interest, to the ion being considered. 
This expression is summed over all ions in the structure that contributes to the EFG tensor at the point of interest. Since the EFG is a two-dimensional tensor with nine elements, each element $V_{ij}$ represents a second derivative with $i, j \subset X, Y, Z$.   The EFG tensor is then diagonalized to obtain the principal components $V_{xx}, V_{yy}$, and $V_{zz}$, where, by convention, $|V_{\rm xx}|< |V_{\rm yy}|< |V_{\rm zz}|$. The principal components are then used to calculate the observables, the asymmetry parameter 
$\eta  \equiv\left |  {V_{xx}-V_{yy}} \right |/{V_{zz}}$  and the splitting $\delta_{q} = \frac{1}{2 h} (eQ)(V_{zz})$, where $eQ$ is the nuclear quadrupole moment. 

\begin{center}	
 %
%%%%%%%%%%%%%%%%%%%%%%%%%%%%%%%%%%%%%%
\begin{figure}[t]
  \vspace*{-0.0cm}
\begin{minipage}{0.98\hsize}
%%%%%%%%%%%%%%%%%%% F I G U R E 6%%%%%%%%%%%%%%%%%%%%
 \centerline{\includegraphics[scale=0.85]{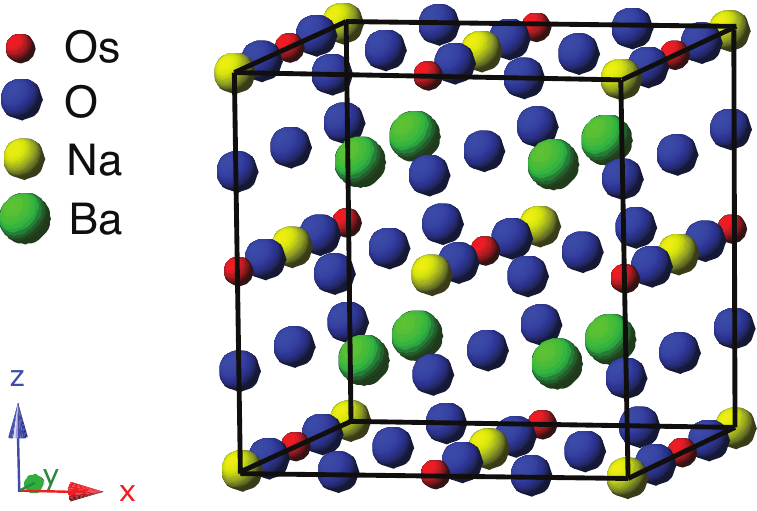}} %%%%%%%%%%%%%%%
%%%%%%%%%%%%%%%%%%%%%%%%%%%%%%%%
\begin{minipage}{.98\hsize}
 \vspace*{-0.0cm}
\caption[]{\label{Cryst} \small %(Color online)  
{\bf a)} Unit cell of Ba$_2$NaOsO$_6$. Oxygen, osmium and sodium ions form face centered cubic structure, while barium ions arrange  a  simple cubic structure. 
This  undistorted double-perovskite structure  has  $Fm\bar3 m$ space group. 
}
 \vspace*{-0.3cm}
\end{minipage}
\end{minipage}
\end{figure}
%%%%%%%%%%%%%%%%%%%%%%%%%%%%%%%%%%%%%%
%
\end{center}
  \vspace*{-0.20cm}
We can then calculate  the EFG tensor at any nuclear site by numerically summing over the lattice for any known   crystal structure.   
The crystal structure of \BaOsS is shown in \mbox{Fig. \ref{Cryst}}.  
At room temperature this material has an undistorted double-perovskite structure, in which OsO$_6$ octahedra are neither distorted nor rotated with respect to each other or the underlying lattice.  
Specifically, considering the periodic nature of crystal structure, one  determines the lattice in a standard way by translation of the three   primary vectors, as described in detail in 
\mbox{Appendix   \ref{Lattice}}. 
 Since  EFG elements are proportional to $1/r^3$, we found that an iteration over 64 unit cells  suffices to make numerical results converge.  That is, summing beyond  64 unit cells does not induce variations in the mean value of any observable that   exceed 1\%. 
In each unit cell, there are 89 ions and  the position of each ion can be accessed by Eq. \ref{position}. 

The first step of the numerical calculation is to set up the distortion model. 
We emphasize that we cannot distinguish between displacements of the actual ions and distortions of the ion charge density in our measurements.  Moreover, the point-charge method does not permit   modeling  distortions of the ion charge density. Therefore, we choose to model the development of the final EFG by local distortions only.  
The local distortion of oxygen ions in an NaO$_6$ octahedron is reflected by altering the basis indices. For example,  in order to test the effect of 2\%'s elongation along $c$-axis of the oxygen ion above the Na site at origin, its original position $(0, 0, 0.25a)$ should be modified to $(0, 0, 0.25(1+2\%)a)$, where $a$ is the lattice constant. After laying out  the distortion model of a single unit cell, we iterate over all unit cells by changing primary vectors. In each unit cell, we first access the position and charge of each ion and then calculate all the EFG elements. Finally, after the iterations, we calculate the asymmetry  parameter $\eta \equiv\left |  {V_{xx}-V_{yy}} \right |/{V_{zz}}$ and maximum $\delta_{q} \equiv \frac{1}{2 h} (eQ)(V_{zz})$ (for $H\|  c$), as derived in  \mbox{Eq. \ref{dq_calApp}}.
 Moreover, even though the calculation can yield $V_{zz}$ parallel to either the $a$ or $b$ crystal 
axes for orthorhombic distortions, our measurements presented here do not allow us to discern between these two orientations of $V_{zz}$. Therefore, we use symmetry arguments to compare results of calculations to our measurements.

\section{Results and Discussion}

Next, we describe results of our calculations of the EFG  induced by various type of distortions  using the point charge approximation. 
 We  first consider distortions where the oxygen ions,  forming the O$^{2-}$ octahedra surrounding Na$^{+}$ ions, are constrained to move along the cubic axes of the perovskite reference unit cell. 

\subsection{Distortion along principal axis - one  structurally distinct Na site}

\begingroup
     \squeezetable
\begin{table}[b]   \centering
  \begin{tabular}{@{} ccccccccc @{}}
\hline $\delta_a$ & $\delta_b$ & $\delta_c$ & $\eta$ & $v_{aa}$ & $v_{bb}$ & $v_{cc}$ & $\delta_q$(kHz) & $V_{zz}$\\\hline
0.1\% & 0.55\% & 0.25\% & 0.87 & -2.39 &  38.89 & -36.49 & 189.8 & $ {b} \,  (y)$\\
\bf 0.55\% & \bf 0.1\% & \bf 0.2\% & \bf 0.87 & \bf 38.89 &  \bf -2.395 & \bf -36.49 & \bf 189.8 & $\bf {a} \, (x)$\\
-0.25\% & -0.65\% & -0.1\% & 0.87 & 2.394 &  -38.89 & 36.49 & 189.8 & ${c}\, (z)$\\
\bf -0.65\% & \bf -0.25\% & \bf -0.1\% & \bf 0.87 & \bf -38.89 &  \bf 2.394 & \bf 36.49 & \bf 189.8 & $\bf  {a}\, (x)$\\\hline
\end{tabular}
\caption[Sample results of point charge calculation with two structurally distinct Na sites.]{Sample results of point charge calculations with one structurally Na site. Program loops through $\delta_a$, $\delta_b$ and $\delta_c$ values within the range of (-5\%, 5\%) and returns combinations of parameters that can yield $\eta$ in the vicinity of 0.87 and $\delta_q \approx 190$ kHz. The parameters that reproduce our experimental findings are in bold fonts. (Model A)}
\label{TblModA}
\end{table}
  \endgroup

\begin{center}	
 %
%%%%%%%%%%%%%%%%%%%%%%%%%%%%%%%%%%%%%%
\begin{figure}[t]
  \vspace*{-0.0cm}
\begin{minipage}{0.98\hsize}
%%%%%%%%%%%%%%%%%%% F I G U R E 7 %%%%%%%%%%%%%%%%%%%%
 \centerline{\includegraphics[scale=0.39]{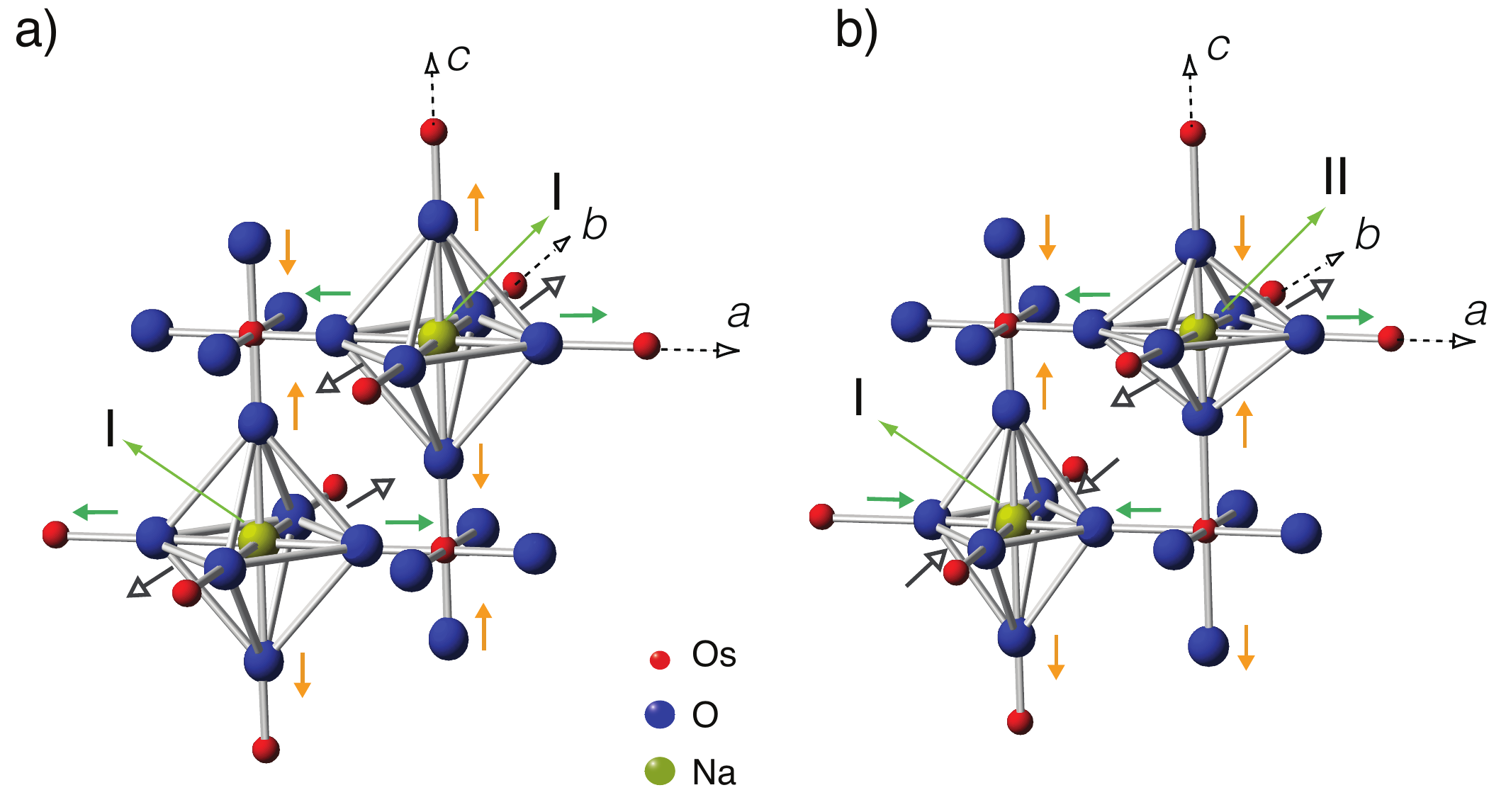}} %%%%%%%%%%%%%%%
%%%%%%%%%%%%%%%%%%%%%%%%%%%%%%%%
\begin{minipage}{.98\hsize}
 \vspace*{-0.0cm}
\caption[]{\label{Fig7} \small %(Color online)  
 Schematic of the proposed lattice distortions. {\bf a)}   One structurally distinct Na site in non-cubic environment  is produced by  elongation/compression of  O$^{2-}$ octahedra. (Model A)
  {\bf b)} 
 Two structurally distinct Na sites are generated by elongation, or compression, of one O$^{2-}$ octahedron  along [001] direction and its concurrent compression, or elongation, in  the (a,b) plane. (Model B)
}
 \vspace*{-0.3cm}
\end{minipage}
\end{minipage}
\end{figure}
%%%%%%%%%%%%%%%%%%%%%%%%%%%%%%%%%%%%%%
%
\end{center}

This model involves distortions of the O$^{2-}$ octahedra surrounding Na$^{+}$ ions  as illustrated \mbox{Fig. \ref{Fig7}a)}. 
 We assume that the modifications are identical for all the octahedra   and that 
   three pairs of O ions independently move along each crystalline axis. That is, each pair of O ions can either elongate or compress symmetrically about the central Na site by an arbitrary amount along the Na-O bond direction.  We note that we are not making any prior assumptions about the structural symmetry.  The schematic of this model is shown in \mbox{Fig. \ref{Fig7}a)}.  
In the actual simulation  we define the distortion in percentage relative to the Na-O distance, \mbox{2.274 \AA}, of undistorted bond. We also define the   elongation deformation as   negative   and the compression deformation as positive. 

 %and the only constraints are from the experimentally observed \mbox{$\delta_q = 190$ kHz} and \mbox{$\eta%$ = 0.87}.

  \vspace*{-0.90cm}

The simulation is ran to produce combinations of distortions along all three axes of the original  cubic axes of the perovskite reference unit cell which can reproduce our observations. 
Thus, our parameter space consists of three numbers, $\delta_a, \delta_b$ and $\delta_c$, corresponding to distortions along crystalline  $a, b$ and $c$ axes, respectively. We find that    numerous combinations result in  the desired/observed values of $\delta_q$ and $\eta$,  some of which are listed in \mbox{Tab. \ref{TblModA}}.  
  $v_{aa}$, $v_{bb}$ and $v_{cc}$ are the EFG components along the $a, b$ and $c$ axis of the lattice coordinate, and  $V_{zz}$ is  the largest absolute value  of the three by definition. 

As evident in \mbox{Tab. \ref{TblModA}}, orthorhombic distortions with $\eta \approx 0.87$  can induce the desired   value of $\delta_q$ for different values of relative displacement along any of the 3 crystalline axes.  
In addition to the appropriate  value of $\delta_q$, the calculations have to identify a set of relative displacements that generate the EFG with its principal component along $a$-axis to account for the data. 
 The set of displacements that account for our experimental observations are presented in   bold font in  \mbox{Tab. \ref{TblModA}}.  We find that distortion along any particular direction that does not exceed 0.8\% of the respective lattice constant reproduces the EFG parameters,  in agreement with our observations. 
 
\subsection{Distortion along principal axis - two structurally distinct Na sites}

In this second model, two structurally different Na sites are generated by elongation, or compression, of one O$^{2-}$ octahedron  along the [001] direction and its concurrent compression, or elongation, in the (110) plane, as illustrated in \mbox{Fig. \ref{Fig7}b)}.  

\begin{table}[t]   \centering
  \begin{tabular}{@{} ccccccccc @{}}
\hline $\delta_a$ & $\delta_b$ & $\delta_c$ & $\eta$ & $v_{aa}$ & $v_{bb}$ & $v_{cc}$ & $\delta_q$(kHz) & $V_{zz}$\\\hline
-0.185\% & -0.185\% & 0.185\% & 0 & 27.9 &  27.9 & -55.9 & 190.03 & $ {c}\, (z)$\\
0.185\% & 0.185\% & -0.185\% & 0 & -27.9 &  -27.9 & 55.9 & 190.03 & $ {c}\, (z)$\\\hline
\end{tabular}
\caption{Sample results of point charge calculations with two structurally distinct Na sites. (Model B)}
\label{TblModB}
\end{table}

 This model also naturally accounts for the appearance of two magnetically different Na sites. These two sites appear from the two   distinct frequency  shifts for triplet I and II, even if the magnetically ordered  state is a simple ferromagnet where all  spins on Os$^{7+}$ ions are assumed to point in the same direction. 
 The transfer hyperfine field from Os electronic spins  to the Na nuclei is mediated by O$^{2-}$ ions via its \mbox{$p$-$d$} hybridization  with well localized 5d orbital of Os$^{7+}$. Evidently, the shorter the distance between  O$^{2-}$  and 
Os$^{7+}$ ions, the stronger the hybridization and thus transfer hyperfine field at the Na site. Thus, the internal field at the Na site in the lower plane (in \mbox{Fig. \ref{Fig7}b)}) consist sof a sum of two stronger and four weaker fields, while the field at Na in the upper plane consists 
of a sum of four stronger and two weaker fields. Consequently, NMR signal from  the lower plane Na will appear at smaller absolute frequency shift (as is the case for triplet I), while that from the upper plane at the larger absolute frequency shift (triplet II). 
 
\begin{center}	
 %
%%%%%%%%%%%%%%%%%%%%%%%%%%%%%%%%%%%%%%
\begin{figure}[b]
  \vspace*{-0.0cm}
\begin{minipage}{0.98\hsize}
%%%%%%%%%%%%%%%%%%% F I G U R E 8 %%%%%%%%%%%%%%%%%%%%
 \centerline{\includegraphics[scale=0.27]{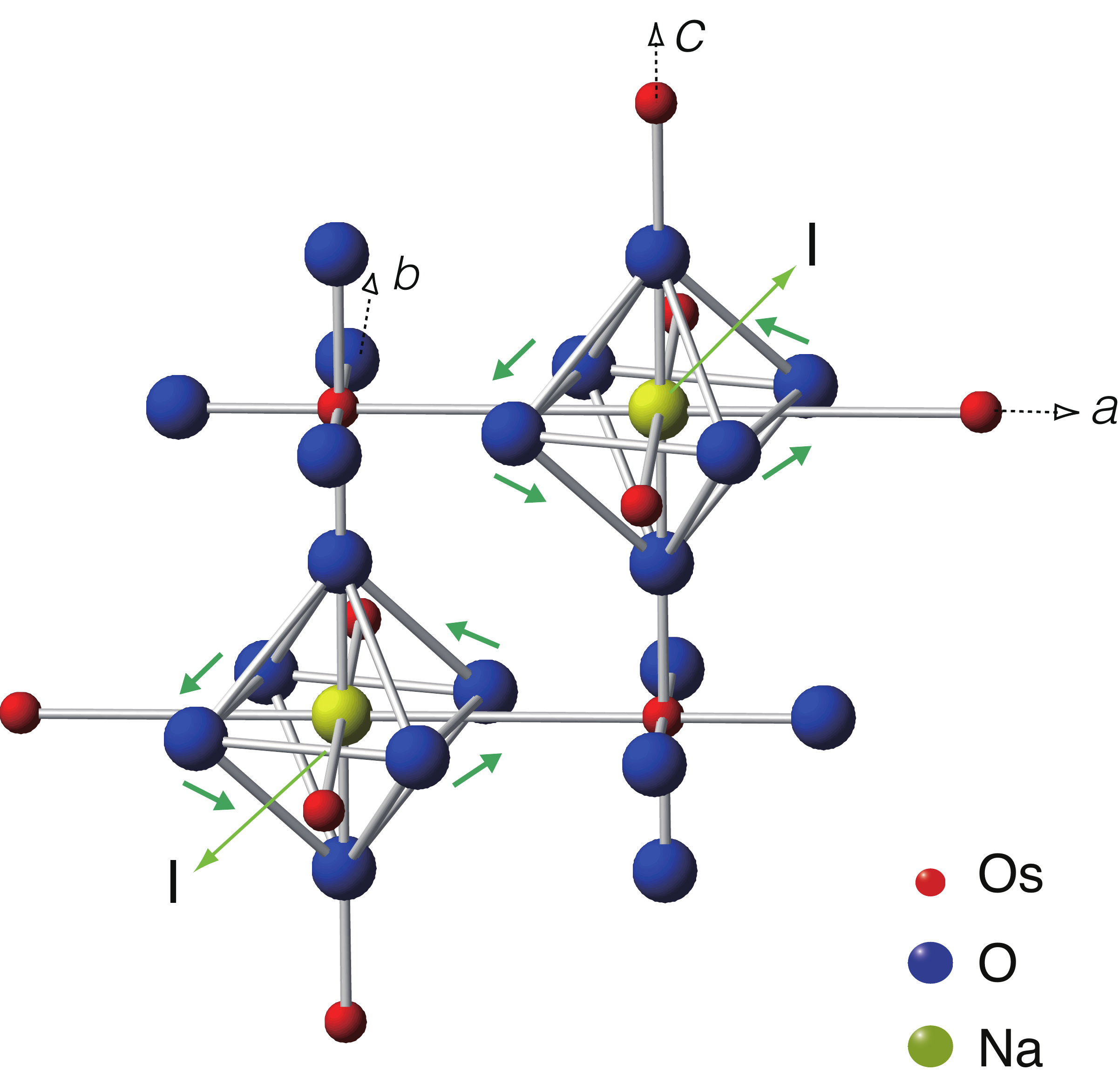}} %%%%%%%%%%%%%%%
%%%%%%%%%%%%%%%%%%%%%%%%%%%%%%%%
\begin{minipage}{.98\hsize}
 \vspace*{-0.0cm}
\caption[]{\label{Fig8} \small %(Color online)  
Schematic of the rotational lattice distortion generating  single Na site. One structurally distinct Na site in non-cubic environment  is produced by in-plane rotation of  O$^{2-}$ octahedra, as depicted by shorter green arrows. (Model C)
}
 \vspace*{-0.3cm}
\end{minipage}
\end{minipage}
\end{figure}
%%%%%%%%%%%%%%%%%%%%%%%%%%%%%%%%%%%%%%
%
\end{center}
  \vspace*{-0.50cm}
  
The findings of our point charge simulations indicate that in order to generate equal $\delta_q$ for both Na sites, as we established in our experiments, the relative magnitude of the elongation has to be equal to that of the compression, \ie distortions must satisfy the relation  $|\delta_a| = |\delta_b| = |\delta_c|$. 
For example, taking the entries from a row of \mbox{Tab. \ref{TblModB}},   (0.185\% , 0.185\% , -0.185\%) distortion, depicted by \mbox{\textbf{I}} in \mbox{Fig. \ref{Fig7}b)}, can reproduce the same $\delta_q$ as the (-0.185\% , -0.185\% , 0.185\%) distortion,  depicted by \mbox{\textbf{II}}. It is very unlikely that such  distortions will occur, as electrostatic energies associated with elongation and compression of the octahedra by the same relative amount are very different. However, as it results from our calculations, such distortions are of  tetragonal symmetry and can only induce $\eta = 0$, in contrast to our observations. For  the reasons listed, this model cannot account for our data.

\subsection{Rotational distortion in $(a,b)$-plane}

Here we consider  model C, consisting of  rotations   of oxygen  ions in the (110) plane.  In this   model  for each Na, four of its surrounding O$^{2-}$ ions  in the $(a,b)$-plane undergo a counter-clockwise rotation, as viewed from the top and depicted in  \mbox{Fig. \ref{Fig8}}.

%%%%%%%%%%%%%%%%%%%%%%%%%%%%%%%%%%%%%%Table 4%%%%%%%%%%%%%%%%%%%%%%%%%%%%%%%%%%%%%%%%%%%
\begin{table}[t]   \centering
  \begin{tabular}{@{} ccccccccc @{}}
\hline  $\phi$ ~& $\delta_c$ ~& $\eta$ ~& $v_{aa}$ ~& $v_{bb}$ ~& $v_{cc}$ ~& $\delta_q$(kHz) ~& $V_{zz}$\\\hline
$5^\circ$ ~& 0\% ~& 0 ~& -0.6675 ~&  -0.6675 ~& 1.335 ~& 4.899 ~& $ {c}$\\
$10^\circ$ ~& 0\% ~& 0 ~& -2.465 ~& -2.465 ~& 4.93 ~& 18.091 ~& $ {c}$\\
$15^\circ$ ~& 0\% ~& 0 ~& -5.19 ~& -5.19 ~& 10.38 ~& 38.091 ~& $ {c}$\\
$20^\circ$ ~& 0\% ~& 0 ~& -8.529 ~& -8.529 ~& 17.06 ~& 62.603 ~& $ {c}$\\
$25^\circ$ ~& 0\% ~& 0 ~& -12.09 ~& -12.09 ~& 24.19 ~& 88.764 ~& $ {c}$\\
$30^\circ$ ~& 0\% ~& 0 ~& -15.46 ~& -15.46 ~& 30.92 ~& 113.46 ~& $ {c}$\\
$35^\circ$ ~& 0\% ~& 0 ~& -18.21 ~& -18.21 ~& 36.43 ~& 133.7 ~& $ {c}$\\
$40^\circ$ ~& 0\% ~& 0 ~& -20.02 ~& -20.02 ~& 40.04 ~& 146.96 ~& $ c$\\
$45^\circ$ ~& 0\% ~& 0 ~& -20.65 ~& -20.65 ~& 41.3 ~& 151.57 ~& $ {c}$\\
$50^\circ$ ~& 0\% ~& 0 ~& -20.02 ~& -20.02 ~& 40.04 ~& 146.96 ~& $ {c}$\\
$55^\circ$ ~& 0\% ~& 0 ~& -18.21 ~& -18.21 ~& 36.43 ~& 133.7 ~& $ {c}$\\
$60^\circ$ ~& 0\% ~& 0 ~& -15.46 ~& -15.46 ~& 30.92 ~& 113.46 ~& $ {c}$\\
$65^\circ$ ~& 0\% ~& 0 ~& -12.09 ~& -12.09 ~& 24.19 ~& 88.764 ~& $ {c}$\\
$70^\circ$ ~& 0\% ~& 0 ~& -8.529 ~& -8.529 ~& 17.06 ~& 62.603 ~& $ {c}$\\
$75^\circ$ ~& 0\% ~& 0 ~& -5.19 ~& -5.19 ~& 10.38 ~& 38.091 ~& $ {c}$\\
$80^\circ$ ~& 0\% ~& 0 ~& -2.465 ~& -2.465 ~& 4.93 ~& 18.091 ~& $ {c}$\\
$85^\circ$ ~& 0\% ~& 0 ~& -0.6675 ~&  -0.6675 ~& 1.335 ~& 4.899 ~& $ {c}$\\\hline
\end{tabular}
\caption{Sample results of point charge calculations of rotational distortion in the (a,b) plane (Model  C). }
\label{TblModC}
\end{table}
%%%%%%%%%%%%%%%%%%%%%%%%%%%%%%%%%%%%%%%%%%%%%%%%%%%%%%%%%%%%%%%%%%%%%%%%%%%%%%%%%%%%%%%%%

%%%%%%%%%%%%%%%%%%%%%%%%%%%%%%%%%%%%%%Table 5%%%%%%%%%%%%%%%%%%%%%%%%%%%%%%%%%%%%%%%%%%%
\begin{table}[b]   \centering
  \begin{tabular}{@{} ccccccccc @{}}
\hline  $\delta_c$ & $\phi$ ~& $\eta$ ~& $v_{aa}$ ~& $v_{bb}$ ~& $v_{cc}$ ~& $\delta_q$(kHz) ~& $V_{zz}$\\\hline
1\% ~& $20^\circ$ ~& 0 ~& 26.18 ~&  26.18 ~& -52.35 ~& 192.13 ~& $ {c}$\\
0.9\% ~& $15^\circ$ ~& 0 ~& 25.8 ~&  25.8 ~& -51.59 ~& 189.34 ~& $ {c}$\\
-0.15\% ~& $45^\circ$ ~& 0 ~& -25.41 ~&  -25.41 ~& 50.82 ~& 186.52 ~& $ {c}$\\
-0.2\% ~& $40^\circ$ ~& 0 ~& -26.35 ~&  -26.35 ~& 52.69 ~& 193.38 ~& ${c}$\\
-0.25\% ~& $35^\circ$ ~& 0 ~& -26.09 ~&  -26.09 ~& 52.18 ~& 191.51 ~& ${c}$\\
-0.35\% ~& $30^\circ$ ~& 0 ~& -26.4 ~&  -26.4 ~& 52.81 ~& 193.8 ~& ${c}$\\
-0.45\% ~& $25^\circ$ ~& 0 ~& -26.06 ~&  -26.06 ~& 52.12 ~& 191.29 ~& ${c}$\\
-0.55\% ~& $20^\circ$ ~& 0 ~& -25.48 ~&  -25.48 ~& 50.95 ~& 187 ~& ${c}$\\
-0.7\% ~& $15^\circ$ ~& 0 ~& -26.52 ~&  -26.52 ~& 53.05 ~& 194.69 ~& ${c}$\\
-0.75\% ~& $10^\circ$ ~& 0 ~& -25.24 ~&  -25.24 ~& 50.48 ~& 185.26 ~& ${c}$\\
-0.85\% ~& $5^\circ$ ~& 0 ~& -26.29 ~&  -26.29 ~& 52.59 ~& 193 ~& ${c}$\\\hline
\end{tabular}
\caption{Sample results of point charge calculations of rotational distortion in the (a,b) plane with  $c$-axis elongation, or compression. (Model C$_{2}$). }
\label{TblModC4a}
\end{table}
%%%%%%%%%%%%%%%%%%%%%%%%%%%%%%%%%%%%%%%%%%%%%%%%%%%%%%%%%%%%%%%%%%%%%%%%%%%%%%%%%%%%%%%%%

The results of point charge simulations are symmetric about [110]  $(\theta = 45^\circ)$, as expected for  the proposed in-plane rotation distortion. We found that our calculations could not generate the splitting \dqS of 190 kHz, as shown in  \mbox{Tab. \ref{TblModC}}.  As a matter of fact, the largest splitting obtained by this model was $\approx 152$~ kHz,   in disagreement with the experiment. Moreover, only tetragonal distortions with $\eta=0$ are generated, which is   inconsistent with our data as well.

     \begin{center}	
 %
%%%%%%%%%%%%%%%%%%%%%%%%%%%%%%%%%%%%%%
\begin{figure}[t]
  \vspace*{-0.0cm}
\begin{minipage}{0.98\hsize}
%%%%%%%%%%%%%%%%%%% F I G U R E 9 %%%%%%%%%%%%%%%%%%%%
 \centerline{\includegraphics[scale=0.33]{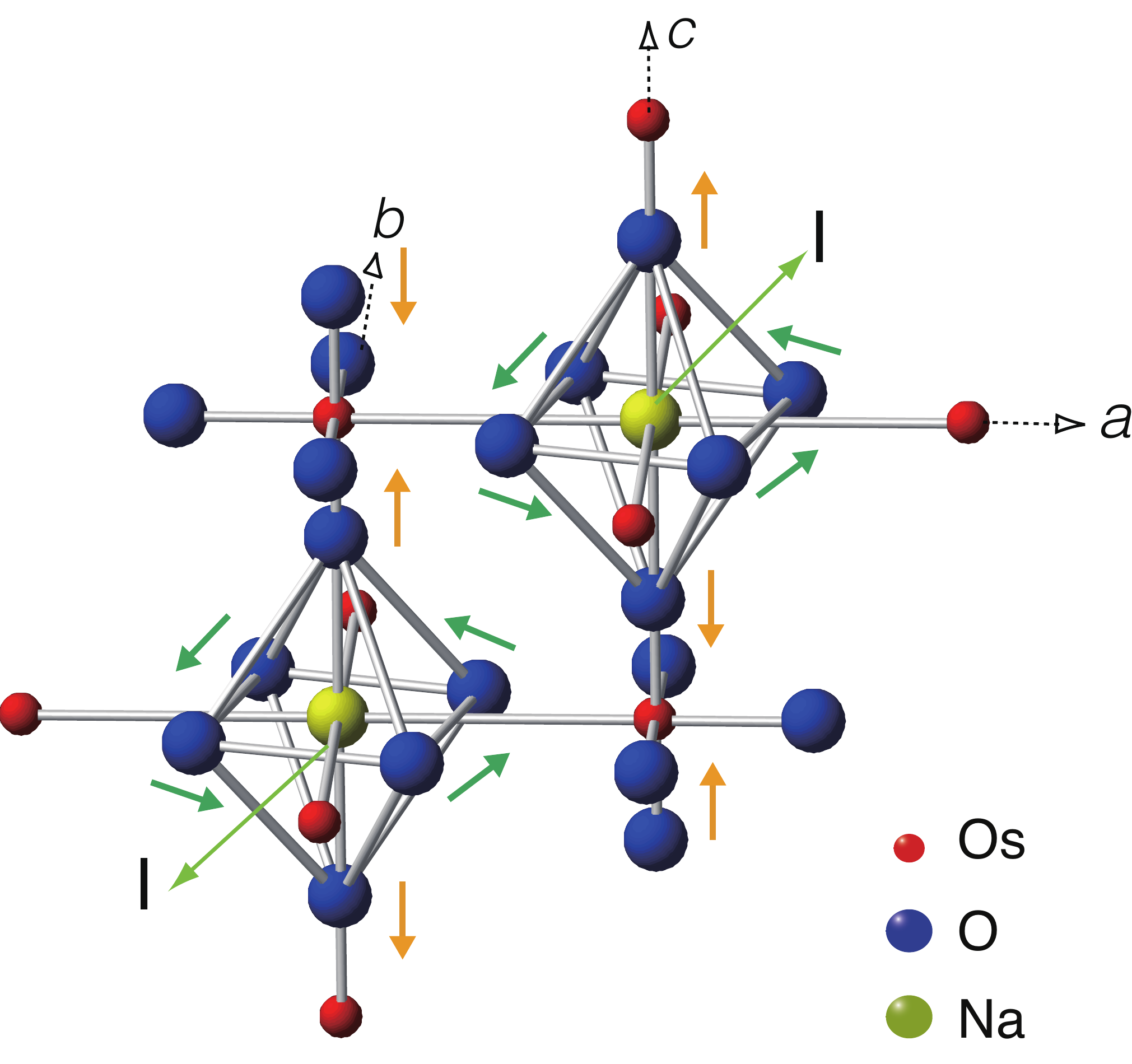}} %%%%%%%%%%%%%%%
%%%%%%%%%%%%%%%%%%%%%%%%%%%%%%%%
\begin{minipage}{.98\hsize}
 \vspace*{-0.0cm}
\caption[]{\label{Fig9} \small %(Color online)  
Schematic of the rotational lattice distortion combined with elongation,  or compression, of two oxygen ions along [001] direction.   One structurally distinct Na site in non-cubic environment  is produced by this lattice modification, consisting of in-plane rotation (green arrows) and distortion along  $c$-axis  (orange arrows) of  O$^{2-}$ octahedra. (Model C$_{2}$)
}
 \vspace*{-0.3cm}
\end{minipage}
\end{minipage}
\end{figure}
%%%%%%%%%%%%%%%%%%%%%%%%%%%%%%%%%%%%%%
%
\end{center}
  \vspace*{-0.90cm}

  Next, we add distortions that involve the other two $O^{2-}$.    Simple arguments indicate that  this type of proposed distortion possesses  tetragonal symmetry, \ie $\eta = 0$. 
  In \mbox{Tab. \ref{TblModC4a}} we display distortions that generate the  observed splitting in point charge simulations.  Positive values in the $\delta_c$ column represent compression while negative ones represent elongation.  
  Calculations   indicate that distortions, comprising from the  elongations along the $c$ axis that  do not exceed 4\% and rotations in the (a,b) plane by the angle defined  with respect to either $a$ or $b$ axis  ranging from  $5^\circ$ to $45^\circ$, can reproduce the measured \dqS of $\approx 190$ kHz. 
   However, this model only generates tetragonal distortions with $\eta = 0$ and $V_{zz} \| c$, both inconsistent with our experimental findings. 
Therefore, both models with dominant rotational distortions in the $(a,b)$ plane fail to account for  our  data.

\begin{center}	
 %
%%%%%%%%%%%%%%%%%%%%%%%%%%%%%%%%%%%%%%
\begin{figure}[t]
  \vspace*{-0.0cm}
\begin{minipage}{0.98\hsize}
%%%%%%%%%%%%%%%%%%% F I G U R E 10 %%%%%%%%%%%%%%%%%%%%
 \centerline{\includegraphics[scale=0.29]{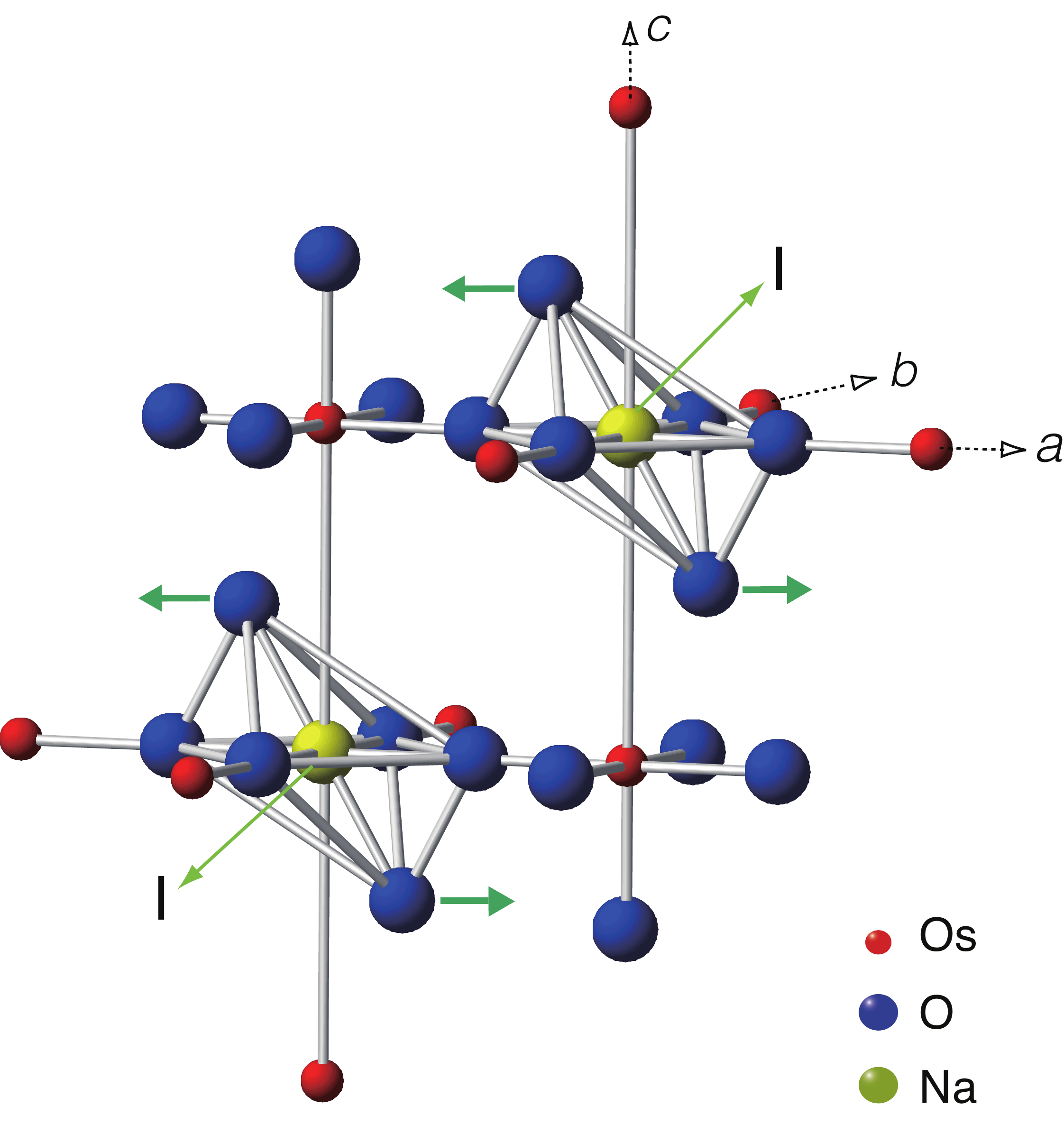}} %%%%%%%%%%%%%%%
%%%%%%%%%%%%%%%%%%%%%%%%%%%%%%%%
\begin{minipage}{.98\hsize}
 \vspace*{-0.0cm}
\caption[]{\label{Fig10} \small %(Color online)  
  Schematic of the tilt-lattice distortion . One structurally distinct Na site in non-cubic environment  is produced by  tilt of the $(a,c)$ plane. (Model D)
}
 \vspace*{-0.3cm}
\end{minipage}
\end{minipage}
\end{figure}
%%%%%%%%%%%%%%%%%%%%%%%%%%%%%%%%%%%%%%
%
\end{center}
  \vspace*{-0.90cm}

%%%%%%%%%%%%%%%%%%%%%%%%%%%%%%%%%%%%%%Table 6%%%%%%%%%%%%%%%%%%%%%%%%%%%%%%%%%%%%%%%%%%%
\begin{table}[b]   \centering
  \begin{tabular}{@{} ccccccccc @{}}
\hline   $\theta$ ~& $\eta$ ~& $v_{aa}$ ~& $v_{bb}$ ~& $v_{cc}$ ~& $\delta_q$(kHz) ~& $V_{zz}$\\\hline
$10^\circ$ ~& 0.6877 ~& -41.46 ~&  6.473 ~& 34.99 ~& 184.67 ~& $ {a}$\\
$10.1^\circ$ ~& 0.6888 ~& -41.97 ~&  6.529 ~& 35.44 ~& 187.03 ~& $ {a}$\\
$10.2^\circ$ ~& 0.6899 ~& -42.48 ~&  6.586 ~& 35.9 ~& 189.41 ~& $ {a}$\\
$10.3^\circ$ ~& 0.691 ~& -43 ~&  6.644 ~& 36.36 ~& 191.82 ~& $ {a}$\\
$10.4^\circ$ ~& 0.692 ~& -43.52 ~&  6.702 ~& 36.82 ~& 194.25 ~& $ {a}$\\
$10.5^\circ$ ~& 0.6931 ~& -44.05 ~&  6.76 ~& 37.29 ~& 196.7 ~& $ {a}$\\\hline
\end{tabular}
\caption{Sample results of point charge calculations of tilt distortion in the $(a,c)$ plane (Model D). }
\label{ModelD}
\end{table}
%%%%%%%%%%%%%%%%%%%%%%%%%%%%%%%%%%%%%%%%%%%%%%%%%%%%%%%%%%%%%%%%%%%%%%%%%%%%%%%%%%%%%%%%%

\subsection{Tilt distortion in $(a,c)$-plane}

In the following model,  we consider tilt distortions in the $(a,c)$ plane depicted in \mbox{Fig. \ref{Fig10}}. 
Two O$^{2-}$  ions of the  octahedra position along  the $c$ axis are tilted by an angle $\theta$ away from the $c$ axis  in the $(a,c)$ plane.  The rest of crystal remains unchanged. To generate the observed splitting, the tilt angle $\theta$ is found to be   $\approx 10^\circ$. However,  the resulting asymmetry parameter is $\approx   0.69$,  as shown in \mbox{Tab. \ref{ModelD}}. This value of $\eta$ is insuficient to account for the experimental data.

 \subsection{Rotational distortion in (a,b) plane and tilt distortion in $(a,c)$ plane}
\begin{center}	
 %
%%%%%%%%%%%%%%%%%%%%%%%%%%%%%%%%%%%%%%
\begin{figure}[t]
  \vspace*{-0.0cm}
\begin{minipage}{0.98\hsize}
%%%%%%%%%%%%%%%%%%% F I G U R E 11 %%%%%%%%%%%%%%%%%%%%
 \centerline{\includegraphics[scale=0.35]{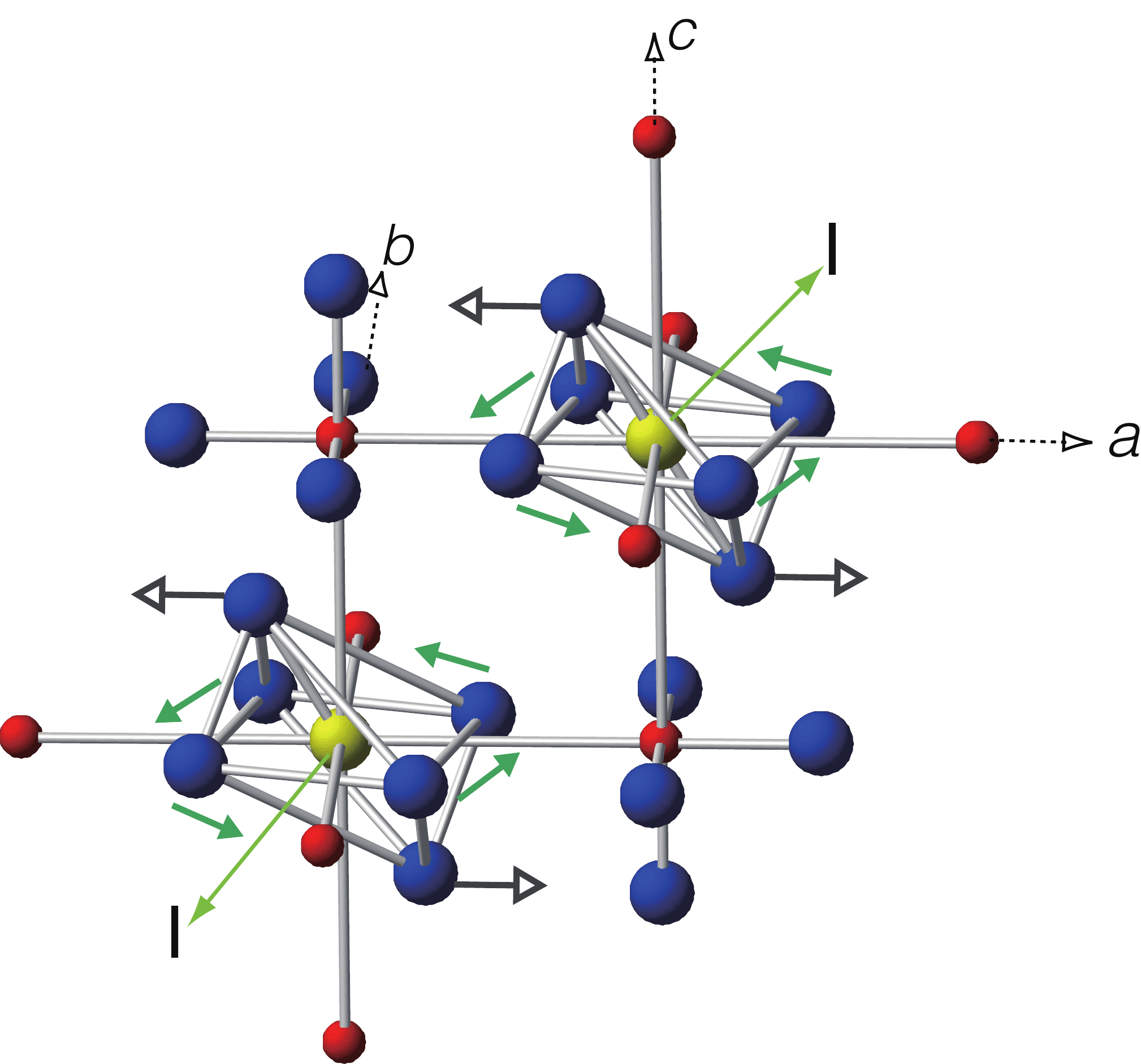}} %%%%%%%%%%%%%%%
%%%%%%%%%%%%%%%%%%%%%%%%%%%%%%%%
\begin{minipage}{.98\hsize}
 \vspace*{-0.0cm}
\caption[]{\label{Fig11} \small %(Color online)  
 Schematic of the combined rotational and tilt lattice distortions. One structurally distinct Na site in non-cubic environment  is induced by concurrent $(a,b)$ plane rotation and $(a, c)$ plane tilt. (Model E)
} 
 \vspace*{-0.3cm}
\end{minipage}
\end{minipage}
\end{figure}
%%%%%%%%%%%%%%%%%%%%%%%%%%%%%%%%%%%%%%
%
\end{center}

In this model,  we consider tilt distortions in  the $(a,c)$ 
%%%%%%%%%%%%%%%%%%%%%%%%%%%%%%%%%%%%%%Table 7%%%%%%%%%%%%%%%%%%%%%%%%%%%%%%%%%%%%%%%%%%%
\begin{table}[h]   \centering
  \begin{tabular}{@{} ccccccccc @{}}
\hline   $\theta$ ~&  $\phi$ ~& $\eta$ ~& $v_{aa}$ ~& $v_{bb}$ ~& $v_{cc}$ ~& $\delta_q$(kHz) ~& $V_{zz}$\\\hline
$6.8^\circ$ ~& $20^\circ$ &  0.8194 ~& -36.26 ~&  -3.6 ~& 39.86 ~& 189.11 ~& $ {c}$\\
$6.9^\circ$ ~& $20^\circ$ &  0.8228 ~& -36.61 ~&  -3.56 ~& 40.17 ~& 190.92 ~& $ {c}$\\
$7.1^\circ$ ~& $19^\circ$ &  0.859 ~& -36.63 ~&  -2.779 ~& 39.41 ~& 190.68 ~& ${c}$\\
$7.3^\circ$ ~& $18^\circ$ &  0.8962 ~& -36.69 ~&  -2.008 ~& 38.69 ~& 190.69 ~& ${c}$\\
$7.4^\circ$ ~& $17^\circ$ &  0.9313 ~& -36.39 ~&  -1.294 ~& 37.69 ~& 189.01 ~& ${c}$\\
$7.5^\circ$ ~& $17^\circ$ &  0.9342 ~& -36.78 ~&  -1.251 ~& 38.03 ~& 190.98 ~& ${c}$\\
$7.6^\circ$ ~& $16^\circ$ &  0.9701 ~& -36.51 ~&  -0.5546 ~& 37.07 ~& 189.52 ~& ${c}$\\
$7.8^\circ$ ~& $15^\circ$ &  0.991 ~& -36.67 ~&  0.165 ~& 36.51 ~& 189.47 ~& ${a}$\\
$8.2^\circ$ ~& $14^\circ$ &  0.9493 ~& -37.7 ~&  0.9555 ~& 36.75 ~& 190.79 ~& ${a}$\\
$8.5^\circ$ ~& $13^\circ$ &  0.9126 ~& -38.41 ~&  1.677 ~& 36.73 ~& 190.83 ~& ${a}$\\
$\bf 8.7^\circ$ ~& $\bf 12^\circ$ &  \bf 0.88 ~& \bf -38.74 ~&  \bf 2.324 ~& \bf 36.41 ~& \bf 189.38 ~& $\bf {a}$\\
$9^\circ$ ~& $11^\circ$ &  0.8489 ~& -39.58 ~&  2.99 ~& 36.59 ~& 190.52 ~& ${a}$\\
$9.2^\circ$ ~& $10^\circ$ &  0.8214 ~& -40.02 ~&  3.573 ~& 36.45 ~& 190.09 ~& ${a}$\\\hline
\end{tabular}
\caption{Sample results of point charge calculations of rotational distortions $(a,b)$ plane and  tilt   in $(a,c)$ plane (Model E). }
\label{tb:pointCharge6}
\end{table}
%%%%%%%%%%%%%%%%%%%%%%%%%%%%%%%%%%%%%%%%%%%%%%%%%%%%%%%%%%%%%%%%%%%%%%%%%%%%%%%%%%%%%%%%%
%
 
%In this model,  we consider tilt distortions in the $(a,c)$ 
\noindent plane accompanied with  a rotational distortion in the $(a,b)$ plane, as described in Model C.  
Specifically, the lattice modification consists of a rotational distortion in $(a,b)$ plane and a tilt in $(a,c)$  plane, as shown in \mbox{Fig. \ref{Fig11}}.  Four O$^{2-}$ ions  of  octahedra positioned in the same plain as Na, \ie 
in $(a,b)$ plane,  are rotated by an angle $\phi$ (counter-clockwise from the $a$ axis).   The  two remaining   O$^{2-}$ ions, located along the $c$ axis,   are tilted by an angle $\theta$  relative to the $c$ axis. 
A subset  of angles $\theta$ and $\phi$   that generate parameters, in point charge approximation, in agreement with experimental observations are listed in \mbox{Tab.\ref{tb:pointCharge6}}. We find that for $\theta = 8.5^{\circ} \pm 0.4^{\circ}$ and $\phi = 12^{\circ} \pm 1^{\circ}$  calculations results are in good agreement with our data. 
Therefore, this is the only model in addition to Model A that well accounts for our observations.

  \vspace*{-0.90cm}

\subsection{Common lattice  distortions in perovskite oxides}

\begin{center}	
 %
%%%%%%%%%%%%%%%%%%%%%%%%%%%%%%%%%%%%%%
\begin{figure}[t]
  \vspace*{-0.0cm}
\begin{minipage}{0.98\hsize}
%%%%%%%%%%%%%%%%%%% F I G U R E 12 %%%%%%%%%%%%%%%%%%%%
 \centerline{\includegraphics[scale=0.31]{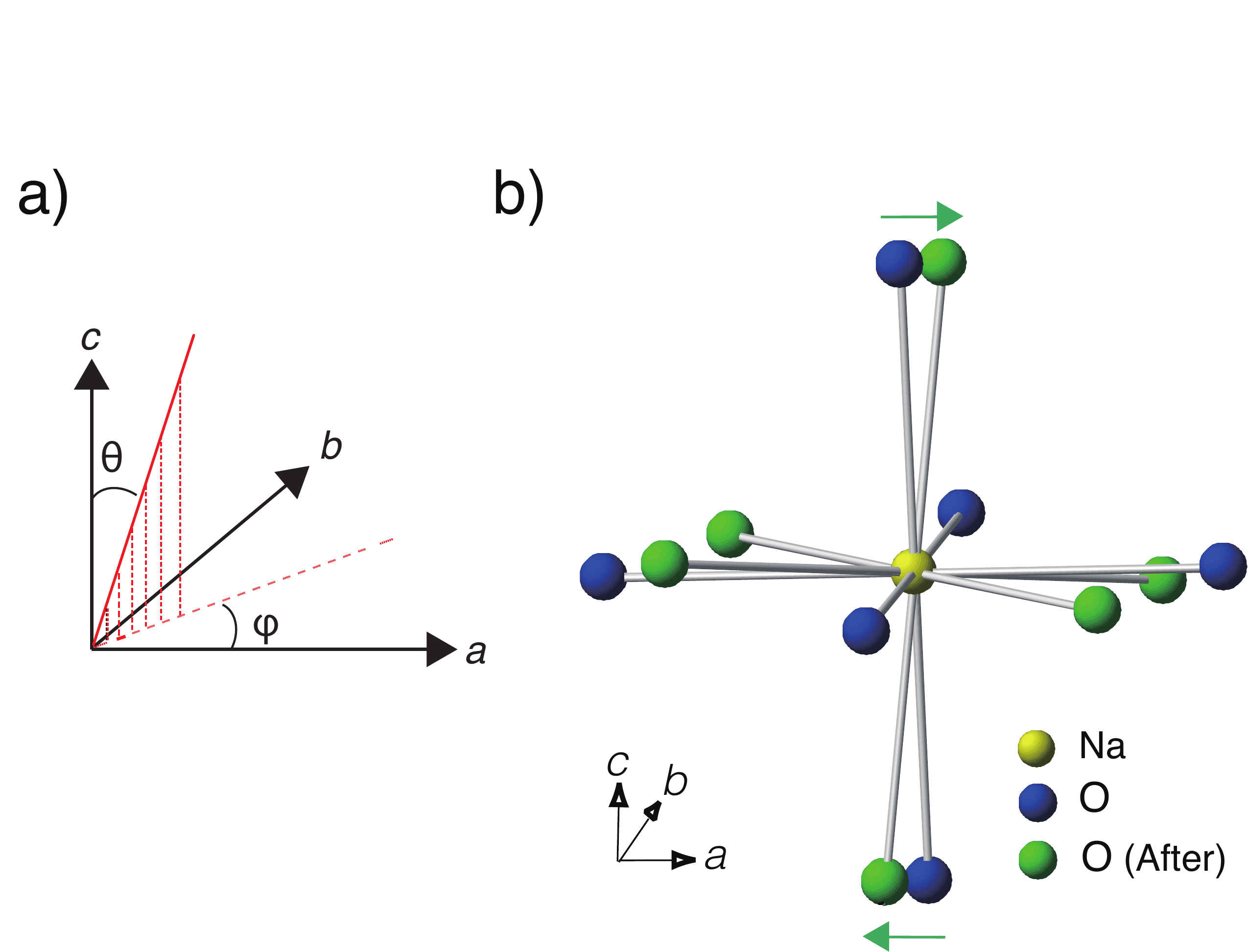}} %%%%%%%%%%%%%%%
%%%%%%%%%%%%%%%%%%%%%%%%%%%%%%%%
\begin{minipage}{.98\hsize}
 \vspace*{-0.0cm}
\caption[]{\label{Fig12} \small %(Color online)  
 Schematic of the proposed GdFeO${_3}$-type  distortion. {\bf a)}  The red solid line denotes distorted Na-O bond, which in its undistorted state points along the   $c$ axis. The NaO${_6}$ octahedra is tilted by angle $\theta$ and $\phi$ in spherical coordinates.  {\bf b)}  Schematic of the specific tilt of NaO${_6}$. The blue spheres/atoms represent O${^{2-}}$ ions in undistorted state while the green ones represent  distorted ones. %$\theta$ is marked. 
 We note that $\phi$ is the angle in $(a,b)$ plane but the  four oxygen ions, originally in $(a,b)$ plane, are no longer in  that plane after the distortion. (Model F)
}
 \vspace*{-0.3cm}
\end{minipage}
\end{minipage}
\end{figure}
%%%%%%%%%%%%%%%%%%%%%%%%%%%%%%%%%%%%%%
%
\end{center}
  \vspace*{-0.40cm}

%%%%%%%%%%%%%%%%%%%%%%%%%%%%%%%%%%%%%%Table 8%%%%%%%%%%%%%%%%%%%%%%%%%%%%%%%%%%%%%%%%%%%
\begin{table}[b]   \centering
  \begin{tabular}{@{} ccccccccc @{}}
\hline   $\theta$ ~&  $\phi$ ~& $\eta$ ~& $v_{aa}$ ~& $v_{bb}$ ~& $v_{cc}$ ~& $\delta_q$(kHz) ~& $V_{zz}$\\\hline
$0^\circ$ ~& $10^\circ$ &  0 ~& -2.465 ~&  -2.465 ~& 4.93 ~& 18.091 ~& ${c}$\\
$0^\circ$ ~& $30^\circ$ &  0 ~& -15.46 ~&  -15.46 ~& 30.92 ~& 113.46 ~& ${c}$\\
$0^\circ$ ~& $44^\circ$ &  0 ~& -20.63 ~&  -20.63 ~& 41.25 ~& 151.39 ~& ${c}$\\
$0^\circ$ ~& $46^\circ$ &  0 ~& -20.63 ~&  -20.63 ~& 41.25 ~& 151.39 ~& ${c}$\\
$0^\circ$ ~& $60^\circ$ &  0 ~& -15.46 ~&  -15.46 ~& 30.92 ~& 113.46 ~& ${c}$\\
$2^\circ$ ~& $10^\circ$ &  0.0721 ~& -2.577 ~&  -2.23 ~& 4.807 ~& 17.687 ~& ${c}$\\
$2^\circ$ ~& $30^\circ$ &  0.0067 ~& -15.49 ~&  -15.29 ~& 30.78 ~& 112.96 ~& ${c}$\\
$2^\circ$ ~& $44^\circ$ &  0.0004 ~& -20.56 ~&  -20.54 ~& 41.1 ~& 150.85 ~& ${c}$\\
$2^\circ$ ~& $46^\circ$ &  0.0004 ~& -20.54 ~&  -20.56 ~& 41.1 ~& 150.85 ~& ${c}$\\
$6^\circ$ ~& $44^\circ$ &  0.0031 ~& -20.08 ~&  -19.96 ~& 40.03 ~& 146.92 ~& ${c}$\\
$6^\circ$ ~& $46^\circ$ &  0.0031 ~& -19.96 ~&  -20.08 ~& 40.03 ~& 146.92 ~& ${c}$\\
$10^\circ$ ~& $44^\circ$ &  0.0086 ~& -19.17 ~&  -18.85 ~& 38.02 ~& 139.55 ~& ${c}$\\
$10^\circ$ ~& $46^\circ$ &  0.0086 ~& -18.85 ~&  -19.17 ~& 38.02 ~& 139.55 ~& ${c}$\\\hline
\end{tabular}
\caption{Sample results of point charge calculations of GdFeO${_6}$ - type distortion with rigid  O${_6}$ octahedra (Model F). %The results are symmetric about $\phi = 45^\circ$ for fixed $\theta$ because $a$ axis and $b$ axis are symmetric about $45^\circ$. 
(Model F)}
\label{tb:pointCharge7}
\end{table}
%%%%%%%%%%%%%%%%%%%%%%%%%%%%%%%%%%%%%%%%%%%%%%%%%%%%%%%%%%%%%%%%%%%%%%%%%%%%%%%%%%%%%%%%%

\begin{center}	
 %
%%%%%%%%%%%%%%%%%%%%%%%%%%%%%%%%%%%%%%
\begin{figure}[t]
  \vspace*{0.1cm}
\begin{minipage}{0.98\hsize}
%%%%%%%%%%%%%%%%%%% F I G U R E 13 %%%%%%%%%%%%%%%%%%%%
 \centerline{\includegraphics[scale=0.30]{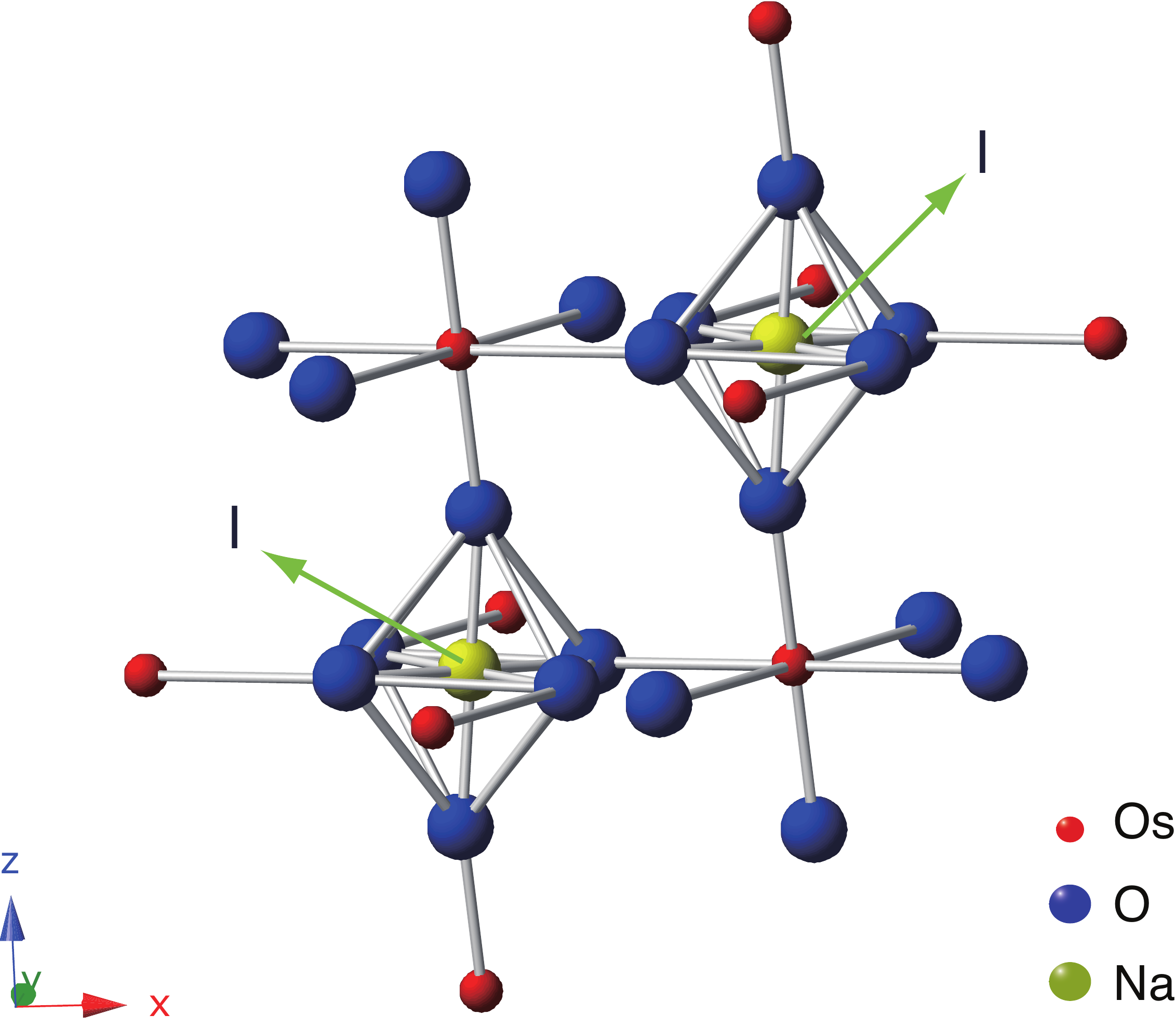}} %%%%%%%%%%%%%%%
%%%%%%%%%%%%%%%%%%%%%%%%%%%%%%%%
\begin{minipage}{.98\hsize}
 \vspace*{-0.0cm}
\caption[]{\label{Fig13} \small %(Color online)  
 Tilting of essentially rigid oxygen octahedra. The schematic of the  tilting of NaO${_6}$ octahedra with $\theta = 10^\circ$ and $\phi = 45^\circ$, where $\theta$ and $\phi$  are standard angles defining spherical coordinates. (Model F$_{2}$)
}
 \vspace*{-0.3cm}
\end{minipage}
\end{minipage}
\end{figure}
%%%%%%%%%%%%%%%%%%%%%%%%%%%%%%%%%%%%%%
%
\end{center}

Perovskite oxides are well known to be prone to lattice distortions \cite{Aleksandrov01, Zhou08}.  
However  the tolerance factor, an indicator   for the stability of crystal structures, 
 of \BaOsS is 0.99, which falls in to the very stable category for cubic structure \cite{Erickson07}.
 Nevertheless,  in this section we consider 
 common lattice instabilities often present in perovskite transition metal oxides, with tolerance factors less than 0.98.   
In general, these lattice distortions involve changes in symmetry and global detectable  changes of lattice parameters.  A typical distortion mechanism  involves a tilting of essentially rigid  oxygen  polyhedra, as is the case in GdFeO$_3$. In \BaOsS this type of modification would tilt the entire rigid oxygen octahedra surrounding  the Na ion. In this case, rigid implies that the octahedra preserve their shape, \ie no motion of individual oxygen atoms occurs.   
However, the orientation of the entire octahedra changes as it tilts away from the [001] direction. 
 The tilt is characterized by two angles: $\theta$, referred to as the polar angle away from $c$ axis, and $\phi$, the azimuthal angle defined relative to the $a$ axis,  as illustrated in \mbox{Fig. \ref{Fig12}a)}.  The tilted crystal structure is depicted in \mbox{Fig. \ref{Fig13}}. The important thing about the GdFeO$_3$-type distortion is that 
 only  the orientation of  the octahedra changes,   whereas their intra-ionic structure remains intact. Moreover, typical values of  the bend in  the bond involving oxygen in GdFeO$_3$ are roughly between 145-170$^{\circ}$.  Point charge calculations are carried for both rigid and non-rigid octahedra cases. 
As it results from our calculations, the maximum quadrupole splitting induced by the distortions involving rigid octahedra does not exceed $\approx 150$ kHz. A subset of the results is displayed in 
\mbox{Tab.\ref{tb:pointCharge7}}. Besides insufficient magnitude of the splitting, in all cases we obtain $V_{{zz}} \| c$, inconsistent with observations.

  \vspace*{-0.90cm}
 
% Point charge calculations are carried for both rigid and non-rigid octahedra case. %The first circumstance assumes that NaO${_6}$ octahedra is rigid (which is more typical) and the second one allows elongation and compression along Na-O bonds.

 %When $\theta$ is fixed, results are symmetric of $\phi = 45$. The reason that calculation results are symmetric about $\phi = 45^\circ$ is $a$ axis and $b$ axis are equivalent and we cannot distinguish them or we can say $a$ axis and $b$axis are symmetric with respect to $\phi = 45^\circ$. Maximum splitting occurs when $\phi = 45^\circ$ for fixed $\theta$. 
%When $\theta$ is fixed, results are symmetric of $\phi = 45$ which is quite obvious due to rotational symmetry. Maximum splitting occurs when $\phi = 45$ for fixed $%\theta$. 
%While $\theta$ increases and $\phi$ is fixed at 45$^\circ$, splitting decreases. Based on point charge calculation, this scenario is not consistent with experimental observations.

%%%%%%%%%%%%%%%%%%%%%%%%%%%%%%%%%%%%%%Table 9%%%%%%%%%%%%%%%%%%%%%%%%%%%%%%%%%%%%%%%%%%%
\begin{table}[b]   \centering
  \begin{tabular}{@{} ccccccccc @{}}
\hline   $\theta$ ~&  $\phi$ ~& $\eta$ ~& $v_{aa}$ ~& $v_{bb}$ ~& $v_{cc}$ ~& $\delta_q$(kHz) ~& $V_{zz}$\\\hline
$\bf 0^\circ$ ~& $\bf 0^\circ$ &  \bf 0.8768 ~& \bf 38.89 ~&  \bf -2.395 ~& \bf -36.49 ~& \bf 189.81 ~& $\bf {a}$\\
$\bf 0^\circ$ ~& $\bf 2^\circ$ &  \bf 0.8719 ~& \bf 38.71 ~& \bf -2.479  ~& \bf -36.24 ~& \bf 188.51 ~& $\bf  {a}$\\
$0^\circ$ ~& $10^\circ$ &  0.7814 ~& 35.17 ~& -3.844  ~& -31.32 ~& 163.8 ~& $ {a}$\\
$0^\circ$ ~& $44^\circ$ &  0.2214 ~& -2.675 ~& -4.196  ~& 6.87 ~& 25.825 ~& $ {c}$\\
$0^\circ$ ~& $46^\circ$ &  0.2214 ~& -4.196 ~& -2.675  ~& 6.87 ~& 25.825 ~& $ {c}$\\
$\bf 2^\circ$ ~& $\bf 0^\circ$ &  \bf 0.8694 ~& \bf 39.13 ~& \bf -2.555  ~& \bf -36.58 ~& \bf 190.31 ~& $\bf  {a}$\\
$\bf 2^\circ$ ~& $\bf 2^\circ$ &  \bf 0.8646 ~& \bf 38.96 ~&  \bf -2.638 ~& \bf -36.32 ~& \bf 189 ~& $\bf {a}$\\
$2^\circ$ ~& $44^\circ$ &  0.2274 ~& -2.612 ~& -4.149  ~& 6.761 ~& 25.445 ~& $ c$\\
$2^\circ$ ~& $46^\circ$ &  0.2274 ~& -4.149 ~& -2.612  ~& 6.761 ~& 25.445 ~& $ {c}$\\
$\bf 4^\circ$ ~& $\bf 2^\circ$ &  \bf 0.8452 ~& \bf 39.58 ~& \bf-3.063  ~& \bf-36.52 ~& \bf190.2 ~& $ {\bf a}$\\
$4^\circ$ ~& $44^\circ$ &  0.2447 ~& -2.442 ~& -4.025  ~& 6.468 ~& 24.436 ~& $ {c}$\\
$4^\circ$ ~& $46^\circ$ &  0.2447 ~&  -4.025 ~&  -2.442 ~& 6.468 ~& 24.436 ~& $ {c}$\\\hline

\end{tabular}
\caption{Sample results of point charge calculations of GdFeO${_6}$ - type distortion with flexible NaO${_6}$ octahedra (Model F + Model A).  The underlying distortions along $a,b$ and $c$ axis are 0.55\%, 0.1\% and 0.25\%. The calculated $\eta$ and $\delta_q$ are symmetric with respect to $\phi = 45^\circ$ because the $a$ axis and $b$ axis are symmetric around $\phi = 45^\circ$. (Model F$_{2}$)} 
\label{tb:pointCharge8}
\end{table}
%%%%%%%%%%%%%%%%%%%%%%%%%%%%%%%%%%%%%%%%%%%%%%%%%%%%%%%%%%%%%%%%%%%%%%%%%%%%%%%%%%%%%%%%%

 Next, we consider the flexible octahedra, where the intra oxygen bonds within an octahedra can   either elongate or compress,  like in models A and B. To illustrate this, we take   the second entry of \mbox{Tab.\ref{TblModA}}. This entry corresponds  to  0.55\% elongation along $a$ axis, 0.1\%  elongation along $b$ axis and 0.25\% compression along  $c$ axis, which generates calculated results in agreement with observations.  We then   combine such elongation/compression and the rigid tilt. In Model A, the displacement of Na-O bond occurs along the $a, b$ or $c$ axis. As  the octahedra tilt, the displacements are still along the Na-O bonds but no longer along the principal axes of the crystal. As   in the case of rigid octahedra, $\theta$ and $\phi$ are the polar and azimuthal angles in spherical coordinates as depicted in \mbox{Fig. \ref{Fig12}a)}. A subset of point charge calculation results is shown in \mbox{Tab.\ref{tb:pointCharge8}}.    
 Minimum splitting occurs when $\phi = 45$ for fixed $\theta$. As illustrated in Table \ref{tb:pointCharge8}, we find that $\theta$ and $\phi$  cannot exceed 4$^\circ$ and 
 2$^\circ$ respectively in order 
  to generate a splitting of  190 kHz. To generate results compatible with our observations, we find that the displacements of oxygen ions must be comparable to those that reproduce the data   in Model A  and 
 the angles  do not   exceed 4$^\circ$. Such  a small tilt angle induces a displacement of oxygen ions that is much smaller than the dominant displacement along the cubic axes of the perovskite reference unit cell. For angles  larger  than $4^\circ$, both $\delta_q$ and $\eta$ significantly decrease below the observed value. Larger  angle  displacements  along the cubic axes are requires to obtain desired $\delta_q$ and $\eta$ values.
 in agreement with the data. This indicates that the dominant displacement of oxygen ions along the cubic axes of the perovskite reference unit cell    reproduce  our observations, as was the case  described in Model A. Essentially, this model maps to   Model A. In addition, the tilt angle has to remain relatively small to assure that these distortions  induce and EFG with the principal axes   aligned with those of the crystal, as imposed by the experiment.  
 Therefore, GdFeO$_3$-type tilt distortions are inconsistent with our data.

\section{Conclusion}

We reviewed details of the quadrupole interactions in Ba${_2}$NaOsO${_6}$ in the low temperature phase characterized by local cubic symmetry breaking. We presented measurements of the splitting \dq, frequency difference between the adjacent   quadrupole perturbed  Zeeman energy levels, at 15 T and 8 K   as a function of the direction of the applied magnetic field. The   field was rotated away from the [001] crystalline axis in two different planes of the crystal.  From the analysis of the rotation data in two different planes,  we established that only the orthorhombic distortions are responsible for the local cubic symmetry breaking in the low temperature magnetically ordered phase. 
These distortions  induce an  EFG with principal axes  aligned with those of the crystal, a principal component $V_{{zz}} \| a$,  and an anisotropy parameter of $\eta \approx 0.87$. 

To find the full set of possible distortions that can induce such EFG, we employed 
the point charge model to calculate $V_{{zz}}$ and $\eta$ resulting from different scenarios. 
 This is the simplest model valid for strongly ionic compounds, such as \BaOs. However, it allowed us to quickly scan through a huge parameter space of possible lattice modifications. 
 We found that our experimental observations can be accounted for by distortions 
 of oxygen octahedra surrounding Na ions  dominated by the 
 displacement of oxygen ions along the cubic axes of the perovskite reference unit cell, as described in Model A.  In addition to distortions in Model A, we find that Model E consisting of combined affects of   tilt distortions in the $(a,c)$ plane  with rotational distortions in the $(a,b)$ plane, for angles not exceeding $12^{\circ}$, is consistent with our data. Both  models are characterized by   the dominant displacement of oxygen ions along the cubic axes of the perovskite reference unit cell.  
 
Since we cannot distinguish between displacements of the actual ions and distortions of the ion charge density, the point charge approach  does not allow us to determine the  nature of 
 the  orbital order  possibly responsible for the local cubic symmetry breaking in \BaOs\cite{ChenBalents10}. First principle calculations of the EFG, with the input from our current work, are required to learn more about the nature of the putative orbital order in this compound.

   \section{Acknowledgement}
 The study was supported in part by the the National Science Foundation  DMR-1608760.       The study at the NHMFL was supported by the National Science Foundation under Cooperative Agreement no. DMR95-27035,  the State of Florida, and Brown University. 
   Work at Stanford University was  supported by the DOE, Office of Basic Energy Sciences, under Contract No.  DE-AC02-76SF00515. 
   
   \section{Appendix}
      \subsection{Quadrupole Interaction}
      \label{QuadApp}
        \subsubsection{Axially Symmetric Case} 
        \vspace*{-0.2cm}
        
   In the simplest case of a field with axial symmetry, the interaction between  the EFG $(eq)$ and the nucleus  with spin $I$ and   quadrupole moment $eQ$,  is described by the  Quadrupole Hamiltonian,
\begin{equation} 
\mathcal{H}_{\rm Q} = \frac{(eQ)(eq)}{4I(2I-1)}[3I_{\rm z}^2-I(I+1)].
\end{equation}
By definition the EFG   is a $3 \times 3$ tensor that corresponds to the rate of change of the electric field at an atomic nucleus. \cite{Kaufmann1979} The matrix is symmetric and traceless. The principal components are denoted by $V_{xx}$, $V_{yy}$, $V_{zz}$ and $|V_{zz}| \geq |V_{yy}| \geq |V_{xx}|$. By convention, the principal component of the EFG is defined as $V_{zz} \equiv eq$.  The principal axes of the EFG define
   the coordinate system $O_{XYZ}$, which is not necessarily aligned with that defined by the crystallographic axes $O_{xyz}$. Evidently, $V_{zz}$ is parallel to one of the crystal axes if 
the principal axes of the EFG and those of the crystal are aligned.

For a nuclear spin $I =3/2$, as is the case of $^{23}$Na, the energy eigenstates of $\mathcal{H}_{\rm Q}$ are given by,
\begin{equation} 
E_{m} =   \frac{(eQ)(eq)}{4I(2I-1)}\, [3m^2-I(I+1)]. 
\end{equation} 
 Then, the frequencies   between different quadrupole satellite transitions equal,
\begin{equation}
\begin{split}
\omega_{\rm m \rightarrow m-1}  = \,  \,   &  \frac{(eQ)(eq)} {h \, 4I(2I-1)} \,[3(2m-1)]  = \frac{(eQ)(eq)} {h} \times \Omega\\ \\
 \Omega \equiv   \frac{1}{2}  ,  & \quad {\rm for} \quad |+ 3/2 \rangle  \rightarrow  |+1/2 \rangle\\
 0   ,&  \quad {\rm for} \quad |+ 1/2 \rangle  \rightarrow  |-1/2 \rangle\\
 -\frac{ 1}{2}   ,&\quad {\rm for} \quad |- 1/2 \rangle  \rightarrow  |-3/2 \rangle\, .\\
\end{split}
\end{equation}

Therefore, in a magnetic field applied along the principal axis of the EFG only 3 NMR lines (transitions) will be observed with equal splitting $\delta_{q}$ between adjacent transitions. In this case, the quadrupole splitting $\delta_{q}$ between different quadrupole satellites is simply given by,  
\begin{equation}
\begin{split}
\label{dq_calApp}
\delta_{q} &= \frac{1}{2 h} (eQ)(V_{zz})\\
&= \frac{1}{2h} \text{(Quadrupole moment)} \times \text{(EFG)}. 
\end{split}
\end{equation}

Consequently, we can estimate the value of the EFG by using the 
experimentally determined value of the splitting   $\delta_{q} \approx 190 \, \rm{kHz}$ for $H \| [001]$.

\begin{equation}
\begin{split}
{\rm EFG} &= \frac{2h\delta_{q}}{eQ} \\
&= \frac{2 \times 4.136\times10^{-15}\, {\rm eV} \cdot s \times 190 \times 10^3\, {\rm s}^{-1}}{0.12 \times e \times 10^{-28}\,{\rm m}^2} \\
&= 1.31 \times 10^{20}\, {\rm V}/{\rm m}^2  .
\end{split}
\end{equation}

Next, this value can be used to roughly estimate the magnitude of particular lattice distortions in our material.  
 In the oxygen octahedra surrounding the Na nuclei the EFG takes on the following form \cite{Choh03},
\begin{equation}
\begin{split}
{\rm EFG} =  
\frac{2q}{4\pi \epsilon _0} 
\left[\begin{matrix}
\begin{smallmatrix}
      \frac{2}{a^3} - \frac{1}{b^3} -\frac{1}{c^3} & 0 & 0\\
      0 & -\frac{1}{a^3} + \frac{2}{b^3} - \frac{1}{c^3} & 0\\
      0 & 0 & -\frac{1}{a^3} - \frac{1}{b^3}+\frac{2}{c^3}\\  
\end{smallmatrix}   
   \end{matrix}\right]
\end{split}
\end{equation}
\vspace*{+0.2cm}

Clearly, with cubic symmetry  such as in Ba$_2$NaOsO$_6$,  the paramagnetic state
is characterized by $a = b=c$.  The EFG   is then   zero, which leads to a vanishing splitting $\delta_{q}$.

The observed $\delta_{\rm q}$ is   largest for a  field applied in the  [001] direction, as shown in Fig. 2 of the manuscript. In this case the simplest model,  accounting for the splitting of the Na line into three equally spaced quadrupole satellite lines,   involves distortions of the O 
 octahedra surrounding Na nuclei solely along the [001] direction. 
In this case, $q = 2e$, $a = b  \neq  c$, and we obtain
\begin{equation}
{\rm EFG} =  \frac{2q}{4\pi \epsilon _0}
\left[\begin{smallmatrix} 
      \frac{1}{a^3} - \frac{1}{c^3}  & 0 & 0\\
      0 & \frac{1}{a^3} - \frac{1}{c^3} & 0\\
      0 & 0 & -2(\frac{1}{a^3} - \frac{1}{c^3})\\      
   \end{smallmatrix}\right] \, .
\end{equation}

Therefore, the principal axis of the EFG $(V_{{zz}} \equiv eq)$ is given by,
\begin{equation}
V_{\rm zz} = \pm\frac{8e}{4\pi\epsilon _0} \left ( \frac{1}{a^3} - \frac{1}{c^3} \right)
\end{equation}

\begin{equation}
\begin{split}
\left ( \frac{1}{a^3} - \frac{1}{c^3} \right)    = & \pm \frac{4\pi\epsilon _0}{8e} \,\times \, 1.31 \times 10^{20}\, {\rm V}/{\rm m}^2    \\ = &  \pm \, 0.01137\times 10^{30}\, {\rm m}^{-3}.
\end{split}
\end{equation}
\\
In \BaOsS  with  $a = 2.274$  \AA, distortions along the $c$ crystalline axis of the order of 4 \% can account for the observed $\delta_{q}$, that is  
\begin{equation}
\begin{split}
&\frac{1}{c^3} = \frac{1}{a^3}\pm 0.01137\\
& c = 2.181 {\rm \AA} \quad (-4.1 \%),   \quad \text{for compression}\\
& c = 2.385 {\rm \AA}  \quad (4.9 \%),   \quad \text{for elongation}.
\end{split}
\end{equation}

%***********************************************************************************************
\vspace*{0.4cm}
  \subsubsection{Anisotropic Charge Distribution Case} 
\label{AnisQ}
\vspace*{-0.3cm}
For anisotropic charge distributions, the quadrupole Hamiltonian expressed in the coordinate system defined by the principal axes of the EFG, is given by
\begin{equation}
\begin{split}
\label{ham}
\mathcal{H}_{\rm Q}(x,y) = \frac{eQV_{zz}}{4I(2I-1)} \left[(3\hat{I}^2_{z}-\hat{I}^2)+\eta(\hat{I}^2_{x}-\hat{I}^2_{y}) \right ],
\end{split}
\end{equation}\\
\vspace*{-0.4cm}
 
\noindent where  $\eta \equiv\left |  {V_{xx}-V_{yy}} \right |/{V_{zz}}$ is the asymmetry parameter  and  V$_{\rm xx}$, V$_{\rm yy}$, and V$_{\rm zz}$ are diagonal components of the EFG.
In this case, the splitting between the adjacent transitions  is given by,
\begin{equation}
\delta_{\rm q} = \frac{(eQ)( V_{\rm zz})}{2 h}   \, \left ( 1 +   \frac{\eta^{2}}{3} \right )^{{1/2}} . 
\end{equation}

Thus, the value of  $\delta_{\rm q}$ is dictated by both $ V_{\rm zz}$ and anisotropy parameter $\eta$. In the high field limit, when $\mathcal{H}_{\rm Q}$ is a perturbation to the dominant Zeeman term, the angular dependence of the splitting is given by
\begin{equation}
\delta_{\rm q} = \frac{\nu_{\rm q}}{2} (3\cos^2\theta - 1 + \eta \sin^2\theta \cos2\phi),
\label{qsplit}
\end{equation}
where   $\theta$ is the angle  between the applied field $ H$ and $V_{zz}$,  $\phi$ is the standard azimuthal angle of spherical coordinate system defined by $O_{{XYZ}}$, and  $\nu_Q \equiv \frac{3e^2qQ}{2hI(2I-1)} = \frac{(eQ)V_{{zz}}}{2h}$.    
As in the case of axially symmetric EFG, in the coordinate system defined by the principal axes of the EFG, denoted by $(O_{XYZ})$, only three NMR lines (transitions) will be observed with equal splitting $\delta_{\rm q}$ between any adjacent lines. \\
%
%\vspace*{-0.5cm}
\subsection{Lattice Sum}
 \label{Lattice}
 
 Considering the periodic nature of crystal structure, one  determines the lattice in a standard way by the  translation of the three   primary vectors. We define the target Na site to be  the origin of three dimensional coordinate system.  Any other point in the lattice is  denoted by
\begin{equation}
\label{pvec}
\bm r'=\bm r + \mu_1\bm a_1 + \mu_2\bm a_2 + \mu_3\bm a_3
\end{equation}
where {$\bm a_i$} are primary vectors of the lattice and {$  \mu_i$}   the primary indices of these vectors.
Then each atom within a unit cell can numerically    be located from 
\begin{equation}
\label{basis}
\bm r_i=x_i\bm a_1 + y_i\bm a_2 + z_i\bm a_3,
\end{equation}
where $x_i$, $y_i$ and $z_i$ are fractions between 0 and 1  that represent the position of the $i^{th}$ atom corresponding to the basis origin.
Combining Eq. \ref{pvec} and Eq. \ref{basis}, the position of any particular ion is then given by
\begin{equation}
\label{position}
\bm r_i=(\mu_1+x_i)\bm a_1 + (\mu_2+y_i)\bm a_2 + (\mu_3+z_i)\bm a_3.
\end{equation}\\

%\clearpage
 $^\dag$ Corresponding author V.F.M. (vemi@brown.edu)
\bibliography{Ref_EFG}
\bibliographystyle{unsrt}

\end{document}